\documentclass{osa-article}

\usepackage{subfigure,epsfig,amsfonts,amsmath,graphicx,dcolumn,bm,array,color,hyperref}
\usepackage{amssymb,setspace,empheq,braket,multirow,fixltx2e} 
\usepackage[utf8]{inputenc}
\usepackage[english]{babel}
\usepackage{physics,cuted}

\begin{document}

\title{Efficient ion-photon qubit SWAP gate in realistic ion cavity-QED systems without strong coupling}

\author{Adrien Borne,\authormark{1,2} Tracy E. Northup,\authormark{3} Rainer Blatt,\authormark{3,4} and Barak Dayan,\authormark{1,*}}

\address{\authormark{1}AMOS and Department of Chemical and Biological Physics, Weizmann Institute of Science, Rehovot 76100, Israel\\
\authormark{2}Laboratoire Matériaux et Phénomènes Quantiques, Université Paris Diderot, CNRS, UMR 7162, 75013 Paris, France\\
\authormark{3}Institut für Experimentalphysik, Universität Innsbruck, Technikerstraße 25, 6020 Innsbruck, Austria\\
\authormark{4}Institut für Quantenoptik und Quanteninformation, Österreichische Akademie der Wissenschaften, Technikerstraße 21a, 6020 Innsbruck, Austria}

\email{\authormark{*} barak.dayan@weizmann.ac.il}



\begin{abstract}
We present a scheme for deterministic ion-photon qubit exchange, namely a SWAP gate, based on realistic cavity-QED systems with \textsuperscript{171}Yb\textsuperscript{+}, \textsuperscript{40}Ca\textsuperscript{+} and \textsuperscript{138}Ba\textsuperscript{+} ions. The gate can also serve as a single-photon quantum memory, in which an outgoing photon heralds the successful arrival of the incoming photonic qubit. Although strong coupling, namely having the single-photon Rabi frequency be the fastest rate in the system, is often assumed essential, this gate (similarly to the Duan-Kimble C-phase gate) requires only Purcell enhancement, i.e. high single-atom cooperativity. Accordingly, it does not require small mode volume cavities, which are challenging to incorporate with ions due to the difficulty of trapping them close to dielectric surfaces. Instead, larger cavities, potentially more compatible with the trap apparatus, are sufficient, as long as their numerical aperture is high enough to maintain small mode area at the ion’s position. We define the optimal parameters for the gate’s operation and simulate the expected fidelities and efficiencies, demonstrating that efficient photon-ion qubit exchange, a valuable building block for scalable quantum computation, is practically attainable with current experimental capabilities.
\end{abstract}

\section{Introduction}
\label{sec:intro}

Efficient ion-photon qubit exchange is a vital building-block for the modular scaling-up of ion-based quantum information systems \cite{Kimble2008,Luo2009,Duan2010, Monroe2013,Northup2014}. 
Although heralded schemes \cite{Barrett2004,Riebe2004,Olmschenk2009,kurz2014,Pirandola2015,kurz2016,Wan2019} provide a basis for linking separate systems even with non-deterministic photonic links, there are many experimental efforts towards the attainment of efficient ion-photon interfaces via cavity quantum electrodynamics \cite{Herskind2009,Hunger2010,Stute2012,Sterk2012,Brandstatter2013,Steiner2013,Stute2013,Takahashi2013,Cetina2013,Steiner2014,Casabone2015,Eltony2016,Marquez2016,Ballance2017,Takahashi2020}. Typically, these efforts aim at obtaining the highest possible atom-cavity coupling rate ($g$, also termed single-photon Rabi frequency) in order to reach the strong-coupling regime, i.e. the regime in which $g$ is the fastest rate in the system \cite{Kimble1998}. Since $g$ is inversely proportional to the square root of the cavity mode volume ($g \propto 1/\sqrt{V}$), this implies miniature cavities ($< 1$~mm), which in turn make the stability of the trap very challenging due to the proximity of the ion to dielectric surfaces \cite{Harlander2010,Wang2011}. 

\begin{figure}[t]
\begin{center}
	\includegraphics[scale=0.22]{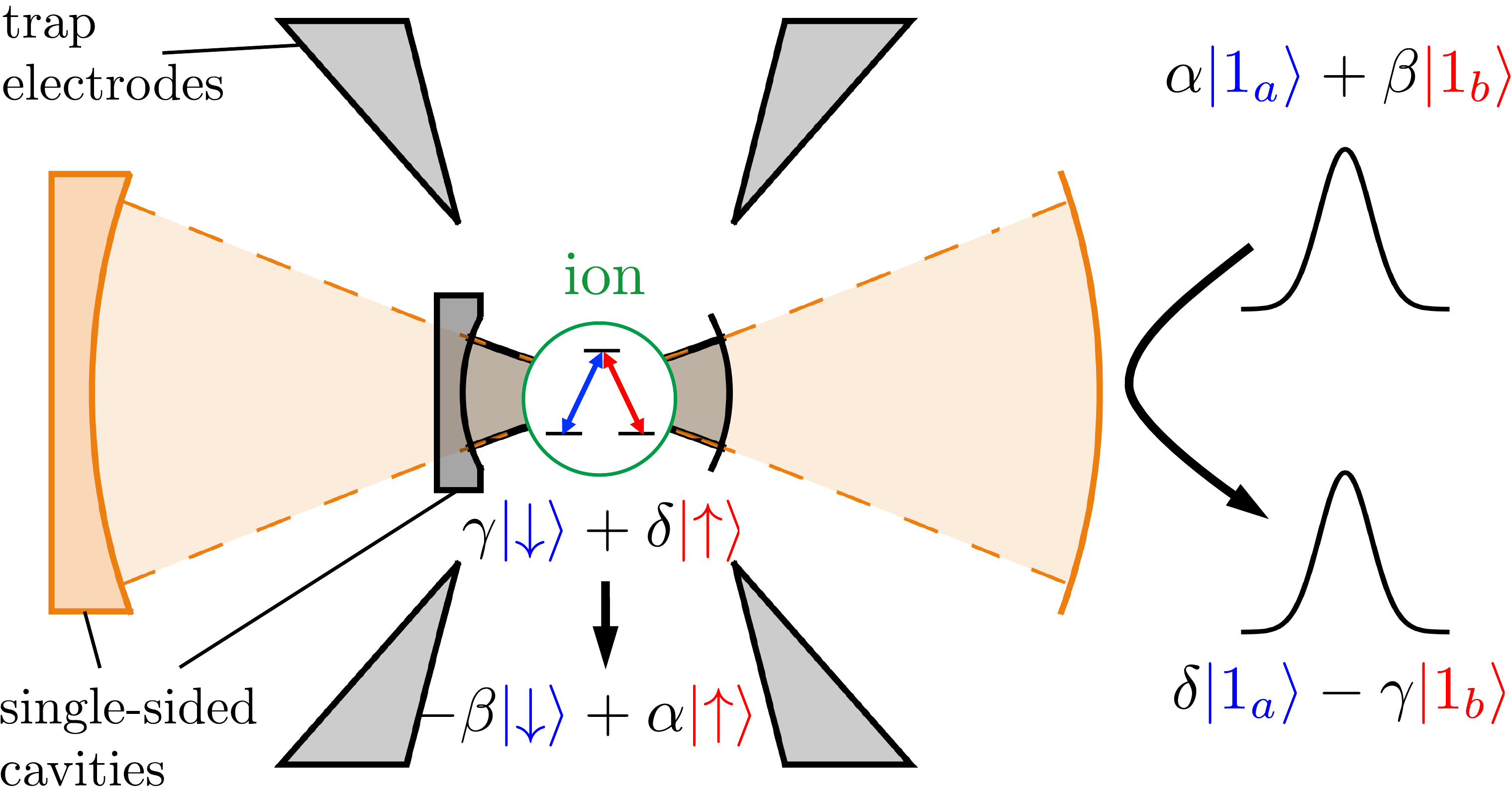}
\end{center}
\vspace{-4mm}
\caption{Two possible cavity configurations for a photon-ion SWAP gate. A photon impinging on an optical single-sided cavity, here represented either as the black or the orange cavity, swaps its state with that of a trapped ion. This optical state is encoded in a superposition of two optical modes corresponding to the two transitions of the ionic $\Lambda$ system. Having the same ratio between the cavity length and the radii of curvature of the mirrors, the two cavity configurations exhibit the same stability properties and the same cooperativity, even though only the smaller one may attain the strong-coupling regime.  
}
\label{fig:concept}
\end{figure}

In contrast, the two native atom-photon gates demonstrated to date, the controlled-phase gate (suggested by Duan and Kimble \cite{Duan2004} and demonstrated experimentally in \cite{Hacker2016} and following works) and SWAP gate (suggested in \cite{Pinotsi2008}, theoretically studied in \cite{Lin2009,Koshino2010,Gea-Banacloche2011,Rosenblum2011,Gea-Banacloche2012,Rosenblum2017} and demonstrated in \cite{Bechler2018}) do not strictly require strong coupling. Both gates do require high cooperativity $C=g^2/\kappa \gamma \gg 1$, where $\kappa$ is the cavity decay rate and $\gamma$ the spontaneous emission rate of the atom into free space. This cooperativity essentially corresponds to Purcell enhancement and is proportional to $Q/V$, with $Q$ being the quality factor of the cavity \cite{Purcell1946}. While this may suggest that small mode volume is required, note that both $Q$ and $V$ scale linearly with the cavity round-trip length $\ell$. This means that the Purcell enhancement - and the cooperativity - do not depend on $\ell$ \cite{Reiserer2015} but are in fact proportional to $F/A$, with $A$ being the mode area and $F$ the finesse of the cavity, namely the cavity lifetime in units of the round-trip time (see Fig.~\ref{fig:concept}). Although this distinction may seem trivial, in ionic systems this is crucial as it allows placing ions sufficiently far from dielectrics. 

In this work we wish to demonstrate quantitatively the feasibility and potential of realizing photon-ion qubit gates in practical systems. We do so by analyzing in detail the implementation of a photon-ion SWAP gate similar to that demonstrated with neutral atoms \cite{Bechler2018}. The underlying mechanism is the single-photon Raman interaction (SPRINT), and the system consists of a three-level $\Lambda$-type quantum emitter inside a single-sided cavity, i.e. a cavity with one of its two mirrors being completely reflective. Both the mechanism and the system are described in the following Sec.~\ref{sec:realization}. 
We then perform an analytical description in Sec.~\ref{sec:model}, which allows us to quantify the performance of the swap process by calculating its fidelity and efficiency for arbitrary ionic and photonic qubits. In Sec.~\ref{sec:ions}, we finally apply our results to realistic situations with \textsuperscript{171}Yb\textsuperscript{+}, \textsuperscript{40}Ca\textsuperscript{+} and \textsuperscript{138}Ba\textsuperscript{+} ions, analytically when the system is invariant under qubit rotation, and numerically otherwise.  
Also, we show that a proper choice of cavity coupling and detuning rates (cavity-field, cavity-ion and potentially Zeeman detunings) leads to optimization of the gate performance. 
We first consider a conventional cm-long Fabry-Perot cavity since we see that high cooperativity is required, but not strong coupling. Although not required, our model is also valid in the case of strong coupling. This led us to consider a fiber-based Fabry-Perot resonator as well in order to display the performances of the gate in that system.

\section{Realization of an ion-photon SWAP gate}
\label{sec:realization} 

\subsection{The underlying mechanism: single-photon Raman interaction}
\label{subsec:SPRINT}

\begin{figure}[t]
\begin{center}
	\includegraphics[scale=0.14]{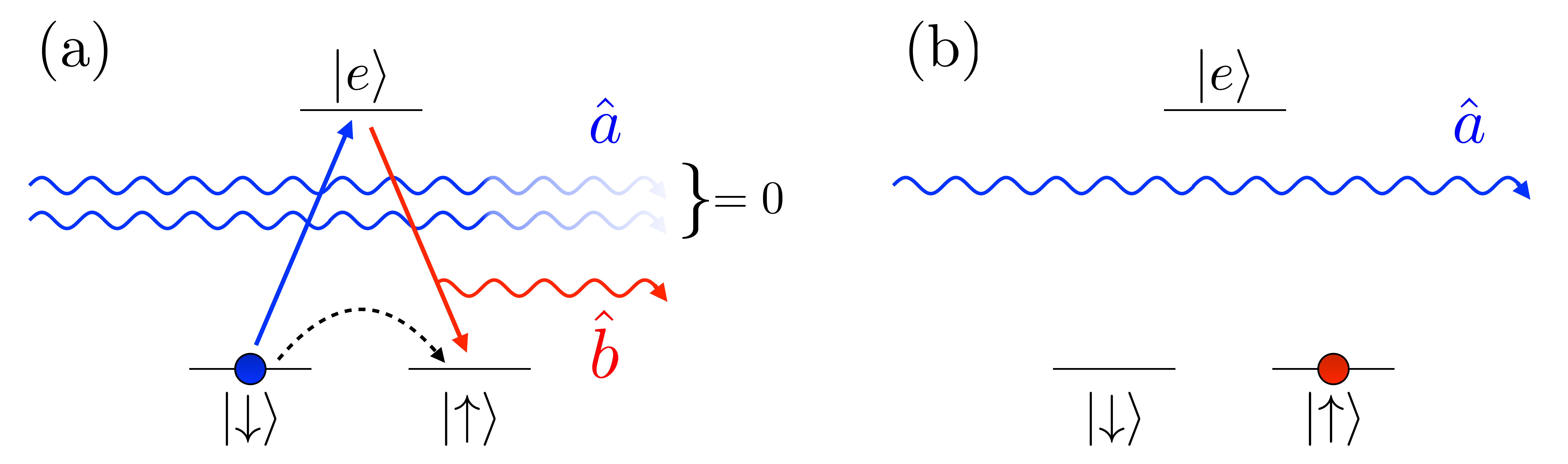}
\end{center}
\vspace{-4mm}
\caption{Operating principle of single-photon Raman interaction (SPRINT). Each transition of a $\Lambda$ system is coupled with high cooperativity to only one mode. (a) Destructive interference between the optical fields in the mode $\hat{a}$ forces the emission of the photon in the mode $\hat{b}$ together with the toggle of the quantum emitter from the state $\ket{\downarrow}$ to $\ket{\uparrow}$. (b) In the toggle, dark state, the photon and the quantum emitter do not interact and hence their states remain unchanged. }
\label{fig:sprint}
\end{figure}

The implementation of the ion-photon SWAP gate under study in this paper relies on a scheme called single-photon Raman interaction (SPRINT) \cite{Pinotsi2008,Lin2009,Koshino2010,witthaut2010,Gea-Banacloche2011,Rosenblum2011,Gea-Banacloche2012,Rosenblum2017}. 
Owing to quantum interference, it allows a single photon to deterministically control the state of a single quantum emitter, and vice versa.
SPRINT requires a three-level $\Lambda$ system to couple independently each of two optical modes ($\hat{a}$ and $\hat{b}$) forming a photonic qubit to one of the two ground states ($\ket{\downarrow}$ and $\ket{\uparrow}$) forming a material qubit; that is, the optical modes $\hat{a}$ and $\hat{b}$ drive the transitions $\ket{\downarrow}-\ket{e}$ and $\ket{\uparrow}-\ket{e}$, respectively. As depicted in Fig.~\ref{fig:sprint}(a) and provided that the two transitions are of equal strength, an incoming probe field of one photon in the mode $\hat{a}$ interacting with the quantum emitter in the state $\ket{\downarrow}$ destructively interferes with the field radiated in the same mode $\hat{a}$, phase-shifted by $\pi$. This leads the system to emit a photon in the mode $\hat{b}$, forcing a Raman transfer of the quantum emitter from $\ket{\downarrow}$ to $\ket{\uparrow}$ \cite{Pinotsi2008,Lin2009,Koshino2010,witthaut2010,Gea-Banacloche2011,Rosenblum2011,Gea-Banacloche2012,Rosenblum2017}. On the other hand, a probe in the mode $\hat{a}$ does not interact with the quantum emitter in $\ket{\uparrow}$, leaving the entire system unchanged as depicted in Fig.~\ref{fig:sprint}(b). 
This nonlinear interaction at the level of a photon is a coherent process that applies also to superposition states of both the photonic and the material qubits and accordingly acts as a SWAP gate between them. Under SPRINT, the photon-emitter joint state is indeed modified as follows: 
\begin{align}
\label{SWAP}
&(c_a \ket{1_a} + c_b \ket{1_b})_{\rm{photon}} \otimes (c_\downarrow \ket{\downarrow} + c_\uparrow \ket{\uparrow})_{\rm{emitter}} \nonumber \\
\longrightarrow \hspace{1mm}
 &(c_\uparrow \ket{1_a} - c_\downarrow \ket{1_b})_{\rm{photon}} \otimes (-c_b \ket{\downarrow} + c_a \ket{\uparrow})_{\rm{emitter}}. 
\end{align}
One of the most trivial applications of the SWAP gate is as a passive quantum memory for a photonic qubit, in which the outgoing photon heralds the successful arrival of the incoming photon. The idea of using the outgoing photon for heralding can be generalised to a sequence of swap operations occurring in various nodes of a quantum network where the outgoing photon from one node can be the input to the next. The outgoing photon of the entire sequence heralds its successful operation. SPRINT can also be used to engineer quantum states of light such as Fock and W states \cite{aqua2019}.

\subsection{Implementation with a single ion trapped in a Fabry-Perot cavity}
\label{subsec:implementation}

This ion-photon SWAP gate can then be implemented by coupling an ion, in which we identify a three-level $\Lambda$ system, to the optical modes through a Fabry-Perot resonator, which provides the necessary interaction enhancement. It has been shown in \cite{Rosenblum2017} that this process can reach unit fidelity by properly choosing the cavity coupling rate so as to get complete destructive interference between the probe and the field radiated in the same mode, provided that the two following conditions are met: (i) the intrinsic loss rate of the resonator must be smaller than the cavity field decay rate, and (ii) the spontaneous emission must be mostly directed into the cavity modes, i.e. $C \gg 1$ as stated in the introduction. 
Note that the coupling rate to the Fabry-Perot resonator is set by the input-output mirror transmission and therefore cannot be fine-tuned, which can reduce the fidelity of the operation. This differs from the case of nanofiber-coupled whispering-gallery mode microresonators (as performed with neutral atoms \cite{Shomroni2014,Rosenblum2016,Bechler2018}), in which coupling rates can continuously be tuned to reach a value optimizing the fidelity. Still, in the case of Fabry-Perot microresonators, we will show in the following that it is possible to circumvent this issue and restore the fidelity by properly setting frequency detuning parameters, at the cost of a reduced efficiency. 
\\

\begin{figure}[t!]
\begin{center}
	\includegraphics[width=0.58\linewidth]{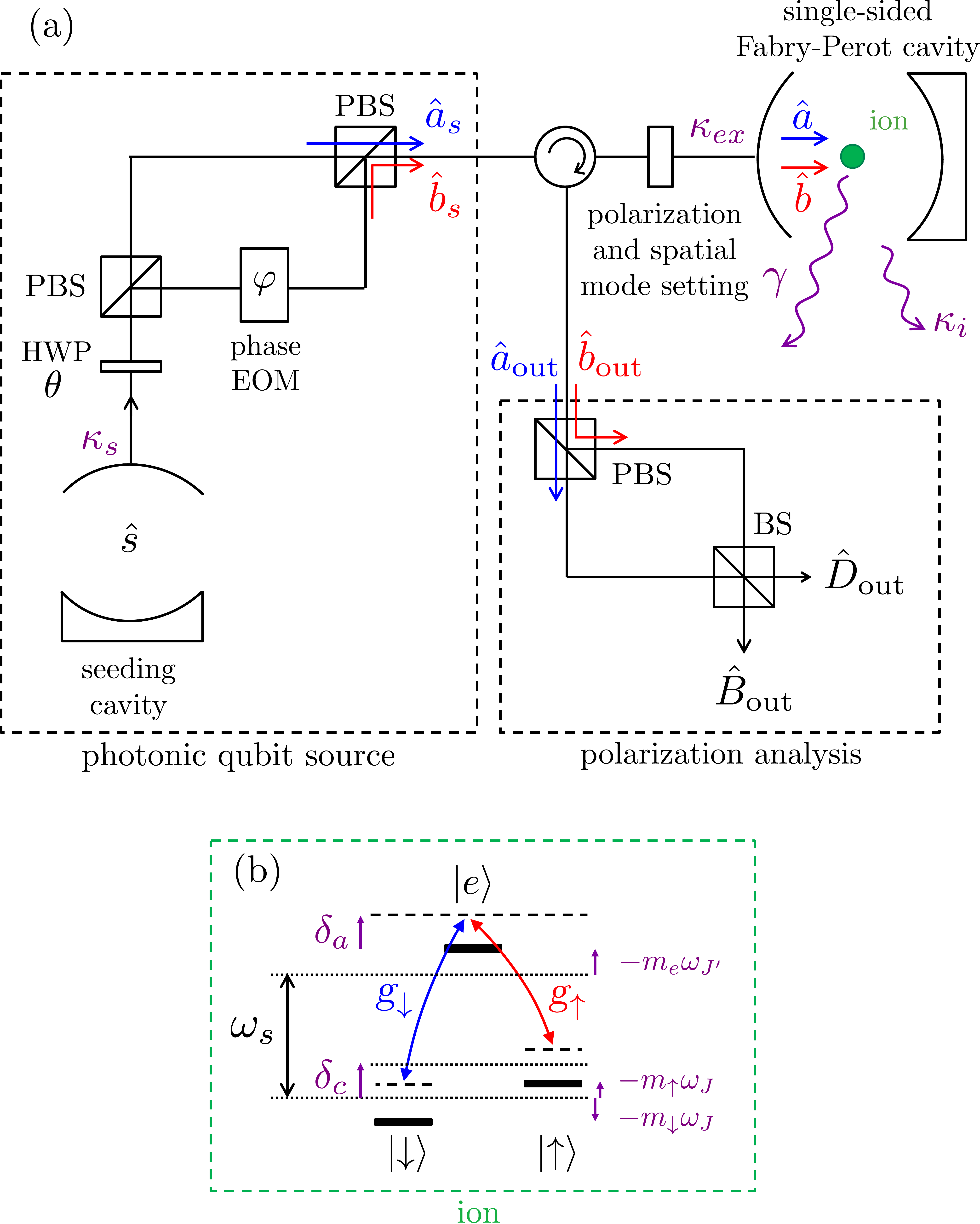}
\end{center}
\vspace*{-4mm}
\caption{Schematic of the theoretical model. (a) The optical setup consisting of a seeding single-sided cavity to generate the photonic qubits and an ion trapped in a second single-sided cavity. (b) The relevant three-level $\Lambda$ system describing the ion. HWP stands for half-wave plate, PBS for polarizing beam splitter and EOM for electro-optic modulator. The various rates and spectral detunings are given in the main text. }
\label{fig:schematics}
\end{figure}

We describe SPRINT in the framework of cascaded systems \cite{Carmichael1993,Rosenblum2011,Rosenblum2017} as shown in Fig.~\ref{fig:schematics}(a). A single-photon input pulse of frequency $\omega_s$ and linearly polarized is modeled by introducing a single-sided seeding cavity described by its annihilation operator $\hat{s}$, which emits with a decay rate $2\kappa_s$. 
Although this restricts the temporal envelope of the single photon pulse to be a decaying exponential, the model will still apply to an arbitrary temporal pulse shape provided that $\kappa_s$ is the lowest rate of the system \cite{Rosenblum2011,Rosenblum2017}. 
A half-wave plate (with angle $\theta/4$ between its fast axis and the incident polarization) and a Mach-Zehnder interferometer composed of two polarizing beam splitters (PBS) with an electro-optic phase modulator (phase $\varphi$) in one arm enable one to define two orthogonal seeding modes of polarization $\hat{a}_s$ and $\hat{b}_s$ such that $\hat{s} = \cos{\left(\theta/2\right)}\hat{a}_s + \sin{\left(\theta/2\right)}e^{i\varphi}\hat{b}_s$. The photon then couples to two orthogonal polarization modes $\hat{a}$ and $\hat{b}$ of a single-sided Fabry-Perot resonator at a rate $2\kappa_{ex}$. Although the second mirror of that single-sided cavity is ideally a perfect reflector, we account for its experimental non-zero transmission, as well as for absorption and scattering, as the intrinsic cavity loss occuring at a rate $2\kappa_i$. The total cavity loss is denoted  $2\kappa_{t}=2(\kappa_{ex}+\kappa_i)$. 

The ion trapped in the resonator can then interact with the photon through the transition from the ground state $\ket{\downarrow}$ (resp. $\ket{\uparrow}$) to the excited state $\ket{e}$, described by the lowering operators $\hat{\sigma}_{\downarrow e}=\dyad{\downarrow}{e}$ (resp. $\hat{\sigma}_{\uparrow e}=\dyad{\uparrow}{e}$), at the rates $2g_\downarrow$ (resp. $2g_\uparrow$). The $\Lambda$-type level scheme is pictured in Fig.~\ref{fig:schematics}(b). We denote the Clebsch-Gordan coefficients associated with these two transitions by $\chi_\downarrow$ and $\chi_\uparrow$, and introduce $g$ such that $g_{\downarrow,\uparrow}=\chi_{\downarrow,\uparrow}g$. The ion can also spontaneously emit in free space at the rate denoted $2\gamma$. 

Using the input-output formalism \cite{Gardiner1985}, we finally define two cavity output modes $\hat{a}_{\mathrm{out}}$ and $\hat{b}_{\mathrm{out}}$ as follows: 
\begin{subequations}
\label{aout,bout}
\begin{align}
\hat{a}_{\mathrm{out}} &= \sqrt{2\kappa_s}\hat{a}_s + \sqrt{2\kappa_{ex}}\hat{a},\\
\hat{b}_{\mathrm{out}} &= \sqrt{2\kappa_s}\hat{b}_s + \sqrt{2\kappa_{ex}}\hat{b}.
\end{align}
\end{subequations}
In order to analyze the polarization of the field exiting the cavity, it is useful to express $\hat{a}_\mathrm{out}$ and $\hat{b}_\mathrm{out}$ in a different basis. This is performed by using a beam splitter, which enables us to define two new output operators $\hat{B}_\mathrm{out}$ and $\hat{D}_\mathrm{out}$ that describe the bright and dark ports of the beam splitter respectively. They will be explicated in the following development (Sec.~\ref{sec:model}).

Since SPRINT relies on destructive interference between the incoming probe and the field radiated by the quantum emitter in the same mode as the probe, a non-perfect spatial overlap of these beams will damage the fidelity of the interaction. For instance, a spatial mode matching of $90\%$ (e.g. in \cite{Gulati2017}) leads to about $5\%$ reduction in fidelity, assuming centered Gaussian beams. It is nonetheless possible to preserve the fidelity by shaping the transverse mode of the probe before the cavity.  
\\

In the following derivation, the system is modeled by considering the general case of an asymmetric $\Lambda$ system, i.e. not invariant under rotation of the Bloch sphere, which can be due to different Clebsch-Gordan coefficients or degeneracy lifting of the Zeeman sublevels. It is therefore here necessary to consider in our description arbitrary input superposition states, unlike in Ref.~\cite{Rosenblum2017}, where the symmetry of the $\Lambda$ system allowed the authors to consider both photonic and atomic states as always residing on the poles of the Bloch spheres. 

\section{Theoretical model}
\label{sec:model} 

The dynamics of the system are given by the following Hamiltonian in a frame rotating at the probe frequency $\omega_s$ ($\hbar=1$):
\begin{subequations}
\label{Hfull}
\begin{align}
\hat{H} =& \hat{H}_{\rm{drive}} + \hat{H}_{\rm{field}} + \hat{H}_{\rm{ion}} + \hat{H}_{\rm{int}}, \\
\hat{H}_{\rm{drive}} =& - 2i\sqrt{\kappa_s\kappa_{ex}}(\hat{a}_s\hat{a}^\dag+\hat{b}_s\hat{b}^\dag)-i\kappa_s \hat{s}^\dag \hat{s}, \\
\hat{H}_{\rm{field}} =& -i[\kappa_{t}+i(\delta_c-m_\downarrow\omega_J)]\hat{a}^\dag \hat{a} \nonumber \\
&-i[\kappa_{t}+i(\delta_c-m_\uparrow\omega_J)] \hat{b}^\dag \hat{b}, \\
\hat{H}_{\rm{ion}} =& -i[\gamma+i(\delta_a-m_e\omega_{J'})]\hat{\sigma}_{ee}, \\
\hat{H}_{\rm{int}} =& (g_\downarrow^\ast\hat{a}^\dag\hat{\sigma}_{\downarrow e}+g_\downarrow\hat{\sigma}_{\downarrow e}^\dag\hat{a})+(g_\uparrow\hat{b}^\dag\hat{\sigma}_{\uparrow e}+g_\uparrow^\ast\hat{\sigma}_{\uparrow e}^\dag\hat{b}), 
\end{align}
\end{subequations}
where $\hat{\sigma}_{ee}=\dyad{e}{e}$ is the population of the excited state. The probe frequency is detuned from the cavity resonance by $\delta_c$, and from the atomic transition by $\delta_a$.  
An applied magnetic field $\vec{B}$ lifts the degeneracy in the total angular momentum and shifts the energy of each of the Zeeman sublevels of magnetic quantum numbers $m_\downarrow$, $m_\uparrow$ and $m_e$. The Larmor frequencies associated with the levels $^{2S+1}L_{J,J'}$ are $\omega_{J,J'}=\mu_B \mathrm{g}_{J,J'} B$ with  $\mu_B$ the Bohr magneton and $\mathrm{g}_{J,J'}$ the Landé factors. 
The non-Hermitian term $\hat{H}_{\rm{drive}}$ accounts for the unidirectional interaction between the seeding cavity and the system. $\hat{H}_{\rm{field}}$ and $\hat{H}_{\rm{ion}}$ are the optical field and ion Hamiltonians, which include losses and detunings, and $\hat{H}_{\rm{int}}$ is the Jaynes-Cummings term describing the photon-ion interaction. 
\\

In the Hilbert space $\mathcal{H} = \mathcal{H}_{s} \otimes \mathcal{H}_{a} \otimes \mathcal{H}_{b} \otimes \mathcal{H}_{\mathrm{ion}}$, where $\mathcal{H}_s$ is associated to the seeding cavity, $\mathcal{H}_a$ and $\mathcal{H}_b$ to the optical modes $\hat{a}$ and $\hat{b}$, and $\mathcal{H}_{\mathrm{ion}}$ to the state of the atom, the initial state can be written: 
\begin{align}
\label{Psi0}
\left|\psi(0)\right\rangle = \left( \alpha \left|1_{a,s} \right\rangle + \beta  \left| 1_{b,s} \right\rangle \right)\otimes \left|0_{a}, 0_{b} \right\rangle  \otimes \left( \alpha' \left| \downarrow \right\rangle + \beta'  \left| \uparrow \right\rangle \right), 
\end{align}
where $|\alpha|^2 + |\beta|^2 = |\alpha'|^2 + |\beta'|^2 =1$, $\alpha = \cos(\theta/2)$ and $\beta = \sin(\theta/2)e^{i\varphi}$. 
According to the Schr\"odinger equation, the state evolves as 
\begin{align}
\label{Psit}
\left|\psi(t)\right\rangle &= 
e^{-\kappa_s t} \left[\alpha\alpha' \left|1_{a,s}, 0_a, 0_b, \downarrow\right\rangle  
+ \alpha\beta' \left|1_{a,s}, 0_a, 0_b, \uparrow\right\rangle \right. \nonumber \\ 
& \left. + \beta\alpha' \left|1_{b,s}, 0_a, 0_b, \downarrow\right\rangle  
+ \beta\beta' \left|1_{a,s}, 0_a, 0_b, \uparrow \right\rangle  \right] \nonumber \\
& + c_1(t)\left|0_{s}, 1_a, 0_b, \downarrow \right\rangle 
+ c_2(t)\left|0_{s}, 1_a, 0_b, \uparrow \right\rangle \\
&+ c_3(t)\left|0_{s}, 0_a, 1_b, \downarrow \right\rangle 
+ c_4(t)\left|0_{s}, 0_a, 1_b, \uparrow \right\rangle \nonumber \\
& + c_5(t)\left|0_{s}, 0_a, 0_b, e \right\rangle, \nonumber
\end{align} 
where
\begin{subequations}
\label{DiffEq}
\begin{align}
\dot{c}_1(t) =& - 2 \alpha\alpha' \sqrt{\kappa_s\kappa_{ex}}e^{-\kappa_s t} -(\kappa_{t} +i\delta_\downarrow) c_1(t) - ig_\downarrow^\ast c_5(t),\\
\dot{c}_2(t) =& - 2 \alpha\beta' \sqrt{\kappa_s\kappa_{ex}}e^{-\kappa_s t} - (\kappa_{t} +i\delta_\downarrow) c_2(t),\\
\dot{c}_3(t) =& - 2 \beta\alpha' \sqrt{\kappa_s\kappa_{ex}}e^{-\kappa_s t} - (\kappa_{t} +i\delta_\uparrow) c_3(t),\\
\dot{c}_4(t) =& - 2 \beta\beta' \sqrt{\kappa_s\kappa_{ex}}e^{-\kappa_s t} - (\kappa_{t} +i\delta_\uparrow) c_4(t) - ig_\uparrow c_5(t),\\
\dot{c}_5(t) =&  - ig_\downarrow c_1(t) - ig_\uparrow^\ast c_4(t) -(\gamma +i\delta_e) c_5(t),
\end{align}
\end{subequations}
and introducing the following notations
\begin{subequations}
\label{dup,ddown,de}
\begin{align}
\delta_q &:= \delta_c-m_q\omega_{J}, 
\end{align}
\end{subequations}
with $q \in \{\downarrow, \uparrow, e \}$. Provided that the driving pulse is long enough such that $\kappa_s$ is the lowest rate of the system, Eq.~(\ref{DiffEq}) are solved by taking $d\ket{\psi(t)}/dt=0 $ at all times. This steady-state solution gives: 
\begin{subequations}
\label{aa,ba,xa}
\begin{align}
c_1(t) & = 2 \frac{\sqrt{\kappa_s\kappa_{ex}}}{\kappa_{t}+i\delta_\downarrow}e^{-\kappa_s t} \times
\left[\frac{2\tilde{C}_{t}}{1+2\tilde{C}_{t}}
\frac{\alpha\alpha' |g_\downarrow|^2 (\kappa_{t}+i\delta_\uparrow) + \beta\beta' g_\downarrow^\ast g_\uparrow^\ast (\kappa_{t}+i\delta_\downarrow)}{|g_\downarrow|^2(\kappa_{t}+i\delta_\uparrow)+|g_\uparrow|^2 (\kappa_{t}+i\delta_\downarrow)} - \alpha\alpha' \right], \\
c_2(t) & = -2\alpha\beta' \frac{\sqrt{\kappa_s\kappa_{ex}}}{\kappa_{t}+i\delta_\downarrow} e^{-\kappa_s t}, \\
c_3(t) & = -2\alpha'\beta \frac{\sqrt{\kappa_s\kappa_{ex}}}{\kappa_{t}+i\delta_\uparrow} e^{-\kappa_s t}, \\
c_4(t) & = 2 \frac{\sqrt{\kappa_s\kappa_{ex}}}{\kappa_{t}+i\delta_\uparrow}e^{-\kappa_s t} \times
\left[\frac{2\tilde{C}_{t}}{1+2\tilde{C}_{t}}
\frac{\alpha\alpha' g_\downarrow g_\uparrow (\kappa_{t}+i\delta_\uparrow) + \beta\beta' |g_\uparrow|^2 (\kappa_{t}+i\delta_\downarrow)}{|g_\downarrow|^2(\kappa_{t}+i\delta_\uparrow)+|g_\uparrow|^2 (\kappa_{t}+i\delta_\downarrow)} - \beta\beta' \right], \\
c_5(t) & = 2i\sqrt{\kappa_s\kappa_{ex}}e^{-\kappa_s t} \frac{2\tilde{C}_{t}}{1+2\tilde{C}_{t}} 
\times \frac{\alpha\alpha' g_\downarrow (\kappa_{t}+i\delta_\uparrow) + \beta\beta' g_\uparrow^\ast (\kappa_{t}+i\delta_\downarrow)}{|g_\downarrow|^2(\kappa_{t}+i\delta_\uparrow)+|g_\uparrow|^2 (\kappa_{t}+i\delta_\downarrow)}, 
\end{align}
\end{subequations}
with $\tilde{C}_{t}$ a complex quantity defined as follows: 
\begin{align}
\label{chitot}
\tilde{C}_{t} = \cfrac{1}{2(\gamma+i\delta_e)}\left(\cfrac{|g_\downarrow|^2}{\kappa_{t}+i\delta_\downarrow} + \cfrac{|g_\uparrow|^2}{\kappa_{t}+i\delta_\uparrow}\right). 
\end{align}
Note that when $\delta_\uparrow=\delta_\downarrow=\delta_e=0$, $\tilde{C}_{t}$ is the total cooperativity $C_{t}$ and quantifies the preferential spontaneous emission of the ion in the cavity modes rather than in free space. 

Then Eqs.~(\ref{aout,bout}) lead us to the output field operators $\hat{a}_{\mathrm{out}}$ and $\hat{b}_{\mathrm{out}}$, and the following beam splitter equations to the dark and bright output field operators $\hat{D}_{\mathrm{out}}$ and $\hat{B}_{\mathrm{out}}$: 
\begin{subequations}
\label{BSeq}
\begin{align}
\hat{D}_{\mathrm{out}} & = \alpha' \hat{a}_{\mathrm{out}} + \beta' \hat{b}_{\mathrm{out}}, \\  
\hat{B}_{\mathrm{out}} & = \beta'^* \hat{a}_{\mathrm{out}} - \alpha'^* \hat{b}_{\mathrm{out}},
\end{align}
\end{subequations}
where $\alpha'$ and $\beta'$, defining the atomic qubit, are also the reflection and transmission coefficients of the beam splitter respectively. It is indeed expected that the outgoing photon carries the initial state of the atom and will exit through the bright port of that fictitious beam splitter. 
In the case of an atomic qubit defined as a pole state of the Bloch sphere, the dark and bright modes simply reduce to the modes $\hat{a}$ and $\hat{b}$. 

Consequently, the probabilities for a single photon to exit the cavity in the dark and bright states are given by:  
\begin{align}
\label{Pdark}
 &  \hspace{-1cm} P_{D}  = \int\limits_0^\infty \langle \hat{D}_{out}^\dag\hat{D}_{out} \rangle\, dt  \\
&  \hspace{-1cm}= \left|\alpha' \left( \alpha\alpha' + \beta\beta' \right)
 + \sqrt{\frac{\kappa_{ex}}{\kappa_s}} \left( \alpha' c_1(0) + \beta' c_3(0)\right)\right|^2 
 +\left|\alpha' \left( \alpha\alpha' + \beta\beta' \right)
 + \sqrt{\frac{\kappa_{ex}}{\kappa_s}} \left( \alpha' c_2(0) + \beta' c_4(0)\right)\right|^2  \nonumber\\ 
& \hspace{-1cm} = \alpha'^2 \left| \alpha\alpha' \left( 1- \frac{2\kappa_{ex}}{\kappa_{t}+i\delta_\downarrow} \right) 
+\beta\beta' \left( 1- \frac{2\kappa_{ex}}{\kappa_{t}+i\delta_\uparrow} \right) 
+ \frac{2\kappa_{ex}}{\kappa_{t}+i\delta_\downarrow} \cfrac{2\tilde{C}_{t}}{1+2\tilde{C}_{t}} \cfrac{\alpha\alpha' |g_\downarrow|^2 (\kappa_{t}+i\delta_\uparrow) + \beta\beta' g_\downarrow^\ast g_\uparrow^\ast (\kappa_{t}+i\delta_\downarrow)}{|g_\downarrow|^2(\kappa_{t}+i\delta_\uparrow) +|g_\uparrow|^2(\kappa_{t}+i\delta_\downarrow)}  \right|^2 \nonumber \\
& \hspace{-1cm} + \beta'^2 \left| \alpha\alpha' \left( 1- \frac{2\kappa_{ex}}{\kappa_{t}+i\delta_\downarrow} \right) 
+\beta\beta' \left( 1- \frac{2\kappa_{ex}}{\kappa_{t}+i\delta_\uparrow} \right) 
+ \frac{2\kappa_{ex}}{\kappa_{t}+i\delta_\uparrow} \cfrac{2\tilde{C}_{t}}{1+2\tilde{C}_{t}} \cfrac{\alpha\alpha' g_\downarrow g_\uparrow (\kappa_{t}+i\delta_\uparrow) + \beta\beta' |g_\uparrow|^2 (\kappa_{t}+i\delta_\downarrow)}{|g_\downarrow|^2(\kappa_{t}+i\delta_\uparrow) +|g_\uparrow|^2(\kappa_{t}+i\delta_\downarrow)}  \right|^2,  \nonumber
\end{align}

\begin{align}
\label{Pbright}
& \hspace{-1cm} P_{B} = \int\limits_0^\infty \langle \hat{B}_{out}^\dag\hat{B}_{out} \rangle\, dt \\
&\hspace{-1cm} = \left|\alpha' \left( \alpha\beta'^* - \beta\alpha'^* \right)
 + \sqrt{\frac{\kappa_{ex}}{\kappa_s}} \left( \beta'^* c_1(0) - \alpha'^* c_3(0)\right)\right|^2 
 +\left|\beta' \left( \alpha\beta'^* - \beta\alpha'^* \right)
 + \sqrt{\frac{\kappa_{ex}}{\kappa_s}} \left( \beta'^* c_2(0) - \alpha'^* c_4(0)\right)\right|^2  \nonumber\\  
&\hspace{-1cm} = \left| \alpha\alpha'\beta'^* \left( 1- \frac{2\kappa_{ex}}{\kappa_{t}+i\delta_\downarrow} \right) 
- \beta\left|\alpha'\right|^2 \left( 1- \frac{2\kappa_{ex}}{\kappa_{t}+i\delta_\uparrow} \right) 
+ \beta'^* \frac{2\kappa_{ex}}{\kappa_{t}+i\delta_\downarrow} \cfrac{2\tilde{C}_{t}}{1+2\tilde{C}_{t}} \cfrac{\alpha\alpha' |g_\downarrow|^2 (\kappa_{t}+i\delta_\uparrow) + \beta\beta' g_\downarrow^\ast g_\uparrow^\ast (\kappa_{t}+i\delta_\downarrow)}{|g_\downarrow|^2(\kappa_{t}+i\delta_\uparrow) +|g_\uparrow|^2(\kappa_{t}+i\delta_\downarrow)}  \right|^2 \nonumber \\
&\hspace{-1cm} + \left| \alpha\left|\beta'\right|^2 \left( 1- \frac{2\kappa_{ex}}{\kappa_{t}+i\delta_\downarrow} \right) 
-\beta\alpha'^*\beta' \left( 1- \frac{2\kappa_{ex}}{\kappa_{t}+i\delta_\uparrow} \right) 
- \alpha'^* \frac{2\kappa_{ex}}{\kappa_{t}+i\delta_\uparrow} \cfrac{2\tilde{C}_{t}}{1+2\tilde{C}_{t}} \cfrac{\alpha\alpha' g_\downarrow g_\uparrow (\kappa_{t}+i\delta_\uparrow) + \beta\beta' |g_\uparrow|^2 (\kappa_{t}+i\delta_\downarrow)}{|g_\downarrow|^2(\kappa_{t}+i\delta_\uparrow) +|g_\uparrow|^2(\kappa_{t}+i\delta_\downarrow)}  \right|^2.  \nonumber
\end{align}
Note that these expressions reduce to that of Ref.~\cite{Rosenblum2017} when the detunings are set to zero and for pole states of the Bloch sphere, e.g. $\alpha=\alpha'=1$. 

We now quantify the operation of the SWAP gate by defining its fidelity $\mathcal{F}$, a figure of merit defined as the overlap of the final state with the expected state. For a given initial state defined by the set of parameters $\{\alpha,\alpha'\}$,
\begin{align}
\label{Fidelity}
\mathcal{F}\left(\alpha,\alpha'\right)=\frac{P_{B}}{P_{B}+P_{D}} 
\end{align}
and $\eta\left(\alpha,\alpha'\right) = P_{B}+P_{D}$ is the efficiency of the process. 
We denote $\overline{\mathcal{F}}$ and $\overline{\eta}$ the average fidelity and efficiency over the initial states.  \\
$P_D$ originates from non-perfect destructive interference between the probe and the field radiated in the same mode as the probe because of intrinsic losses and limited cooperativity. Hence, in order to achieve unit swap fidelity, both the real and the imaginary parts of $P_D$ can be set to zero in Eq.~(\ref{Pdark}). 
We obtain
\begin{subequations}
\label{solutions}
\begin{align}
g_\uparrow &= g_\downarrow^\ast, \\
\omega_J &= \omega_{J'}=0 \Rightarrow \delta_\downarrow = \delta_\uparrow = \delta_c~\mathrm{and} ~ \delta_e=\delta_a.
\end{align}
\end{subequations}
Unit fidelity can then only be reached in symmetric $\Lambda$ systems, whose degenerate transitions have equal Clebsch-Gordan coefficients. 

$\kappa_{ex}$ is set by the transmission of the coupling mirror and is therefore not a good tuning parameter unlike in Ref.~\cite{Rosenblum2017}. However, unit fidelity is found for symmetric $\Lambda$ systems by setting both the probe-cavity and atom-cavity detunings to the following optimal values: 
\begin{subequations}
\label{solutions2}
\begin{align}
\delta_c^{\rm{opt}} &= \pm \left[ \kappa_i \sqrt{4\kappa_{ex}^2(1+C_i)+\kappa_i^2C_i^2}-\kappa_{ex}^2-(1+C_i)\kappa_i^2 \right]^{1/2}, \\
\delta_a^{\rm{opt}} &= \frac{\kappa_i\gamma}{2\delta_c^{\rm{opt}}}\left(2+3C_i -\sqrt{4\frac{\kappa_{ex}^2}{\kappa_i^2}(1+C_i)+C_i^2} \right), 
\end{align}
\end{subequations}
with $C_i = \cfrac{|g_{\downarrow,\uparrow}|^2}{\kappa_i\gamma}$ the intrinsic cooperativity. \\
The parameters $\delta_c^{\rm{opt}}$ and $\delta_a^{\rm{opt}}$ being real, Eqs.~(\ref{solutions2}) only have valid solutions for $\kappa_i \leq \kappa_i \sqrt{1+2C_i} := \kappa_{ex}^{\rm{opt}}$, where $\kappa_{ex}^{\rm{opt}}$ is the optimal coupling to the cavity~\cite{Rosenblum2017}. Specifically, when the impedance-matching condition $\kappa_{ex}=\kappa_{ex}^{\rm{opt}}$ is fulfilled, the system already gives a unit fidelity and no detuning is needed; when $\kappa_{ex}< \kappa_{ex}^{\rm{opt}}$, unit fidelity can be retrieved by tuning the previous detunings according to Eq.~(\ref{solutions2}), at the expense of a decrease of the efficiency; and when $\kappa_{ex}>\kappa_{ex}^{\rm{opt}}$, no correction can be performed here using these detuning parameters. 
It is therefore best to design the experiment by choosing a coupling mirror transmission such that $\kappa_{ex} \overset{<}{\simeq} \kappa_{ex}^{\rm{opt}}$ , the fine-tuning optimization being  performed by setting $\delta_c^{\rm{opt}}$ and $\delta_a^{\rm{opt}}$. 

Regarding the efficiency of the process, it is affected by the intrinsic losses and the spontaneous emission of the ion to free space. Whether the previous optimization of the fidelity is performed or not, the following upper bound holds: 
\begin{align}
\label{max_eta}
\eta \leq \eta_{\rm{max}} = P_B\left(\kappa_{ex}^\mathrm{opt}\right) + P_D\left(\kappa_{ex}^\mathrm{opt}\right) = \left(\cfrac{C_i}{\sqrt{1+2C_i}+1+C_i}\right)^2. 
\end{align} 

We turn next to the case where  $g_\uparrow \neq g_\downarrow$. Although the fidelity will not reach one, its optimization can still be performed numerically by setting $\partial_d F = 0$ with $d\in\{\delta_\downarrow,\delta_\uparrow,\delta_e\}$, i.e.
$ P_{D} .\partial_d P_{B} = P_{B} .\partial_d P_{D}$ averaged over all $ \{ \alpha,\alpha' \} \in [0,1]^2 $ as the system is no longer symmetric under qubit rotations. This procedure then leads to a set of optimal parameters $\{\delta_c^{\rm{opt}},\delta_a^{\rm{opt}},B^{\rm{opt}}\}$. 
Note that it may indeed be helpful in this situation to lift the degeneracy of the Zeeman sublevels by applying the external magnetic field $B^\mathrm{opt}$ in order to compensate for the imbalance in the transition strengths. This way, a larger atom-probe detuning can be chosen for the strongest transition.

\section{Applications to ion systems}
\label{sec:ions}

This section gives two numerical applications of the theoretical model developed in the previous section with actual experimental parameters: the first with a symmetric $\Lambda$ system and the second with an asymmetric $\Lambda$ system. 
In both cases, the quantization axis is chosen along the axis of the cavity to get rid of the $\pi$ transitions (i.e. with no change in magnetic quantum number). If needed, the external magnetic field will be applied along this axis. The optical qubit is then encoded in the two orthogonally polarized circular polarizations ($\hat{L}=\hat{a}$ and $\hat{R}=\hat{b}$) associated with the $\sigma^+$ and $\sigma^-$ transitions. 
Note that if one were to choose a $\Lambda$-system composed of a $\sigma$ and a $\pi$ transitions, the Clebsch-Gordan coefficient associated to the $\sigma$ transition would be multiplied by the geometric factor $1/\sqrt{2}$ to account for the projection of the cavity polarization onto the dipole moment. 

We consider Fabry-Perot resonators that can be conventional macroscopic cavities or fiber-based Fabry-Perot microcavities \cite{Hunger2010}. The latter are composed of laser-machined mirrors at the tip of a fiber, which allows for higher coupling rates but presents higher intrinsic losses. We will show that the SWAP gate protocol is realistic in both cavity systems thanks to Purcell enhancement, regardless of strong coupling. The gate performance indeed won't improve because sub-mm cavities enable one to reach smaller mode volumes. 
Note however that choosing a long cavity, as motivated in our introduction, leads to larger mode diameters on the mirrors, which may limit the finesse of the cavity due to higher sensitivity to surface roughness and curvature imperfections of the mirrors. In the examples below, we therefore considered a cavity length $l = 400\mu$m for fiber-based Fabry-Perot resonators, similarly to \cite{Marquez2016,Gulati2017,Takahashi2020} and about twice as long as in \cite{Steiner2013,Brandstatter2013,Steiner2014,Takahashi2014,Ballance2017} leading therefore to a mode diameter increase of the order of $20\%$. This should prevent any deterioration of the finesse, which has been reported to be significant for $l > 1.5$ mm \cite{Ott2016}. 

We considered in our model the two polarization modes of the cavity to be degenerate in frequency, since situations with no birefringent splitting within the cavity linewidth were observed both in conventional cavities \cite{Stute2013} and fiber-based cavities with mirrors designed with a high degree of rotational symmetry \cite{Takahashi2014}. Nonetheless, our model is valid also when birefringence cannot be neglected. Two different probe-cavity detunings $\delta_{c,h}$ and $\delta_{c,v}$ (the subscripts $h$ and $v$ referring to the horizontal and vertical polarization modes, respectively) therefore need to be introduced in the Hamiltonian Eq.~(\ref{Hfull}), which still remains formally equivalent to the one considered by performing the following change of variables: 
\begin{subequations}
\label{biref}
\begin{align}
\tilde{m}_\downarrow &= m_\downarrow + \cfrac{\delta_{c,v}-\delta_{c,h}}{2\omega_J}, \\
\tilde{m}_\uparrow &= m_\uparrow - \cfrac{\delta_{c,v}-\delta_{c,h}}{2\omega_J}. 
\end{align}
\end{subequations}
Note however that those effective quantities are no longer integers and vary with the external magnetic field. 

\begin{figure}[t]%
\begin{center}
\subfigure{%
\label{fig:levels_Yb_all}%
\includegraphics[scale=0.33]{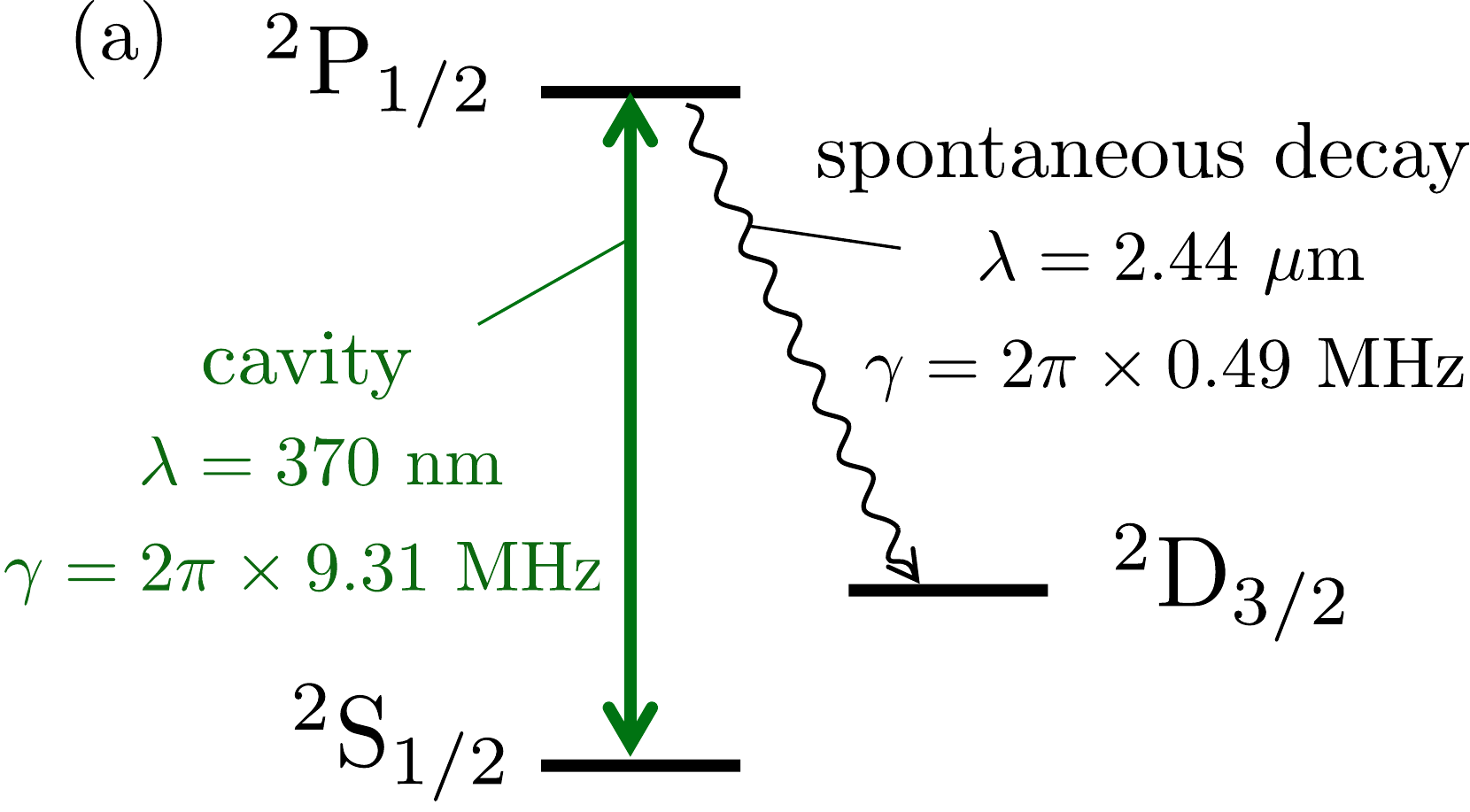}}%
\qquad
\subfigure{%
\label{fig:levels_Yb_lambda}%
\includegraphics[scale=0.33]{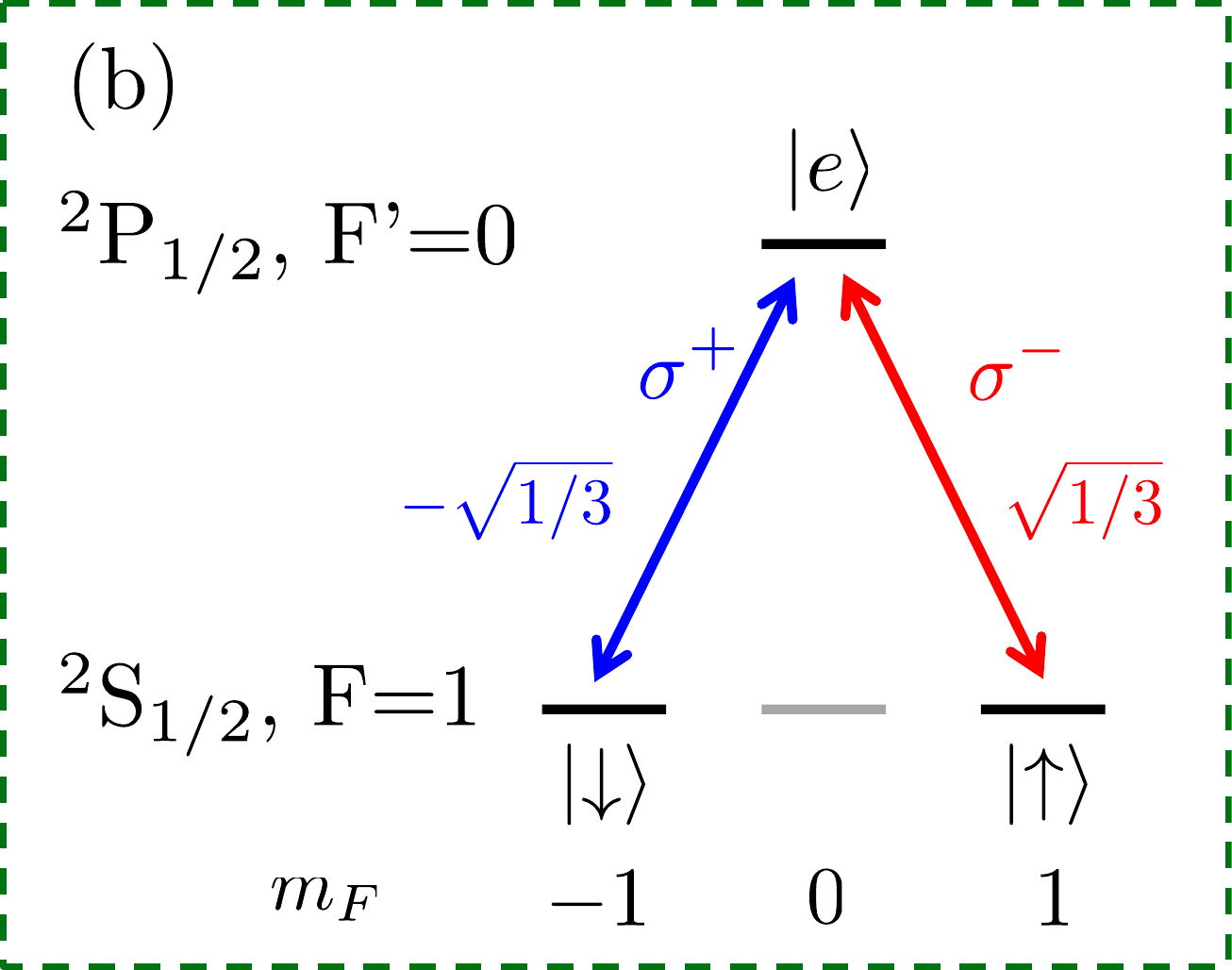}}%
\end{center}
\vspace{-5mm}
\caption{Configuration for a SWAP gate with a symmetric $\Lambda$ system (equal transition strengths). (a) Relevant transitions of \textsuperscript{171}Yb\textsuperscript{+}, and (b) three-level $\Lambda$ system involving the \textsuperscript{2}S\textsubscript{1/2} (F=1) and \textsuperscript{2}P\textsubscript{1/2} (F'=0) manifolds. }
\label{fig:levels_Yb_main}
\end{figure}

\subsection{Symmetric $\Lambda$ system}
\label{subsec:3-1} 
 
We first consider the $^2S_{1/2} - ^2P_{1/2}$ transition, $F=1$ to $F'=0$ in \textsuperscript{171}Yb\textsuperscript{+} at 370 nm, with a spontaneous emission rate into free space $\gamma=2\pi\times 9.8$ MHz (Fig.~\ref{fig:levels_Yb_all}). 
As pictured in Fig.~\ref{fig:levels_Yb_lambda}, $\ket{\downarrow=\{m_F=-1\}}-\ket{e=\{m_{F'}=0\}}$ and $\ket{\uparrow=\{m_F=1\}}-\ket{e=\{m_{F'}=0\}}$ are the two transitions of the $\Lambda$ system, with  strengths of equal magnitude $\chi_\downarrow = -\chi_\uparrow = 1/\sqrt{3} := \chi$. \\

\begin{figure}[t!]%
\begin{center}
\includegraphics[scale=0.55]{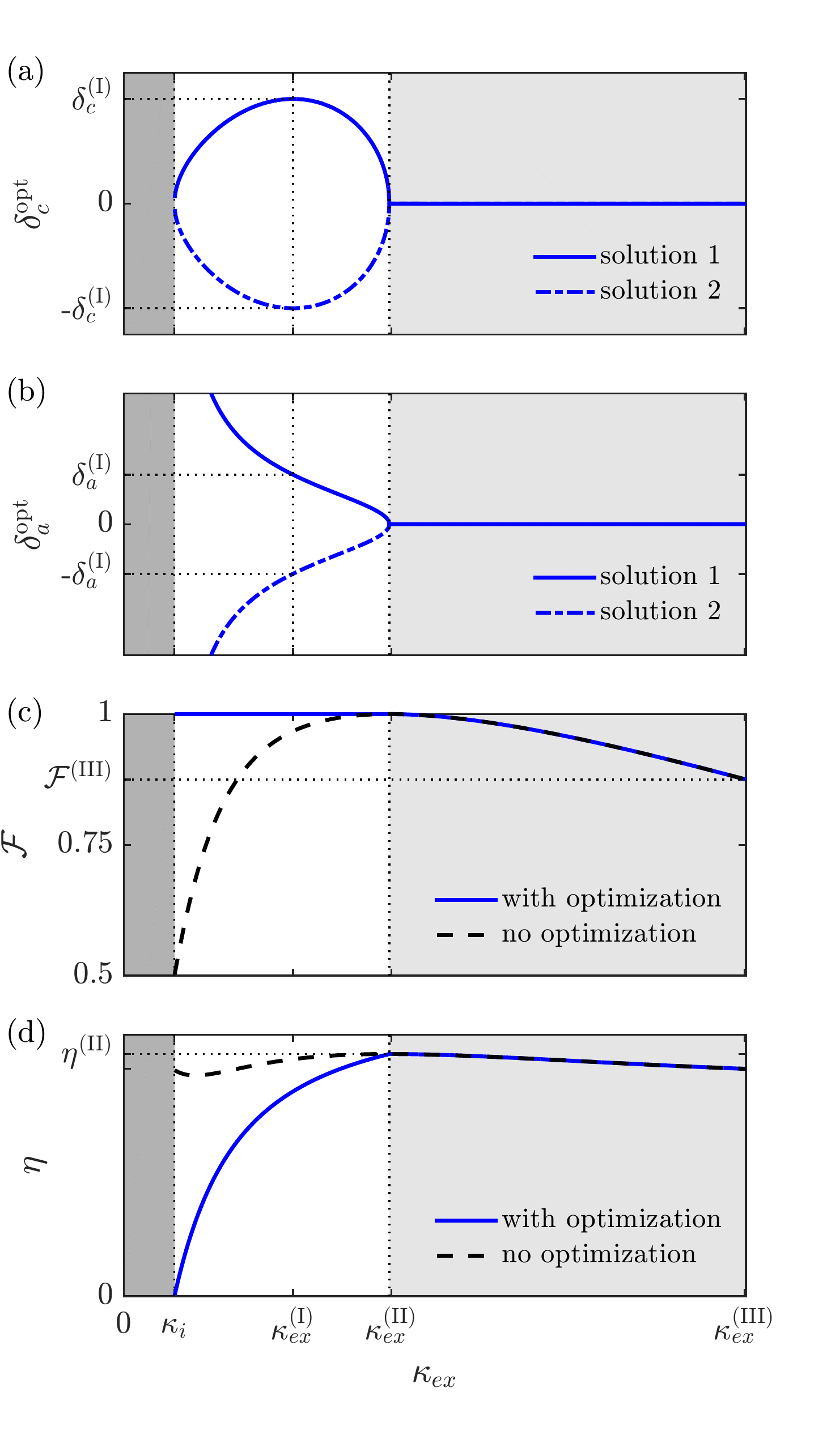} 
\end{center}
\vspace*{-10mm}
\caption{Optimal cavity and atomic detunings $\delta_c^{\rm{opt}}$ (a) and $\delta_a^{\rm{opt}}$ (b) leading to unit swap fidelity within the interval $[ \kappa_i,\kappa_{ex}^{\rm{(II)}}=\kappa_{ex}^{\rm{opt}} ]$ as a function of the extrinsic coupling $\kappa_{ex}$. (c) and (d) shows the optimized (solid blue line) and non-optimized (black dashed line, i.e. with $\delta_c=\delta_a=0$) fidelities and associated efficiencies respectively. Analytical expressions and numerical examples of the various parameters are given in Table~\ref{tab:table}. }
\label{fig:Fig_Sym}%
\end{figure}

\begin{table*}[t!]
\begin{center}
\renewcommand{\arraystretch}{1.5}
    \begin{tabular}{| m{3cm}  | >{\centering\arraybackslash}m{3.5cm}   >{\centering\arraybackslash}m{3cm}   >{\centering\arraybackslash}m{3cm}    }
    \hline
	\hline
    \multicolumn{1}{c}{}    & Analytical expression    & realistic conventional Fabry-Perot cavity   & realistic fiber-based Fabry-Perot cavity  \\ 
	\hline 
    \multicolumn{1}{c}{$\kappa_i$}            &   & 2$\pi\times$ 90 kHz & 2$\pi\times$ 30 MHz \\ 
    \multicolumn{1}{c}{$g_{\downarrow,\uparrow}=\chi g$}       &   & 2$\pi\times$ 2.9 MHz & 2$\pi\times$ 40 MHz \\ 
    \multicolumn{1}{c}{$\gamma$}              &   & 2$\pi\times$ 10 MHz & 2$\pi\times$ 10 MHz \\ 
    \hline 
    \multicolumn{1}{c}{\vspace{1mm} $\kappa_{ex}^{\rm{(I)}}$, for  $\max \delta_c$}   &\vspace{1mm} $\cfrac{\kappa_i}{2} \sqrt{\cfrac{4+8C_i+3C_i^2}{1+C_i}}$  & \vspace{1mm} 2$\pi\times$ 257 kHz & \vspace{1mm} 2$\pi\times$ 69 MHz \\ 
    \multicolumn{1}{c}{$\kappa_{ex}^{\rm{(II)}}=\kappa_{ex}^{\rm{opt}}$}  & $\kappa_i \sqrt{1+2C_i}$  & 2$\pi\times$ 398 kHz &  2$\pi\times$ 103 MHz   \\ 
    \multicolumn{1}{c}{$\kappa_{ex}^{\rm{(III)}}$, for $ C_t=1$} & $\kappa_i(C_i-1)$ & 2$\pi\times$ 743 kHz &   2$\pi\times$ 133 MHz   \\ 
    \hline
    \multicolumn{1}{c}{$\delta_c^{\rm{(I)}}$}      & $\cfrac{\kappa_i C_i}{2\sqrt{1+C_i}}$ & 2$\pi\times$ 130 kHz &     2$\pi\times$ 32 MHz  \\
    \multicolumn{1}{c}{$\delta_a^{\rm{(I)}}$}      & $\gamma\sqrt{1+C_i}$ & 2$\pi\times$ 32 MHz &    2$\pi\times$ 25 MHz   \\
    \multicolumn{1}{c}{$\mathcal{F}^{\rm{(III)}}$} & $\cfrac{4(C_i-1)^2}{20-16C_i+5C_i^2} $  & 0.91  &  0.97   \\
    \multicolumn{1}{c}{$\eta^{\rm{(II)}}=\eta_{\rm{max}}$}         & $\left(\cfrac{C_i}{\sqrt{1+2C_i}+(1+C_i)}\right)^2$ & 0.40 &  0.30   \\
    \multicolumn{1}{c}{$\eta^{\rm{(III)}}$} \vspace{1mm}       & $\cfrac{20-16C_i+5C_i^2}{9C_i^2}$ \vspace{1mm} & 0.39 \vspace{1mm} &  0.30 \vspace{1mm}  \\
    \hline
    \hline
    \end{tabular}
\end{center}
\caption{\label{tab:table} Analytical expressions for the parameters of Fig.~\ref{fig:Fig_Sym}, and numerical values for the 370~nm transition of \textsuperscript{171}Yb\textsuperscript{+} ($^2S_{1/2}, F=1$ to $^2P_{1/2}, F'=0$) for conventional and fiber-based Fabry-Perot resonators.}
\end{table*}

\begin{table*}[h!]
\begin{center}
\renewcommand{\arraystretch}{1.5}
    \begin{tabular}{ >{\centering\arraybackslash}m{2.2cm}   >{\centering\arraybackslash}m{3.5cm}   >{\centering\arraybackslash}m{3cm}   >{\centering\arraybackslash}m{3cm}    }
    \hline
	\hline
    \multicolumn{1}{c}{}    & Analytical expression    & realistic conventional Fabry-Perot cavity    & realistic fiber-based Fabry-Perot cavity  \\ 
	\hline 
    \multicolumn{1}{c}{$\ell$}                &   & 20 mm & 400 $\mu$m \\ 
    \multicolumn{1}{c}{$\mathcal{T}_1$}    &   & 300 ppm & 1500 ppm \\ 
    \multicolumn{1}{c}{$\mathcal{T}_2 + \mathcal{L}$}    &   & 150 ppm & 1000 ppm \\ 
    \hline
    \multicolumn{1}{c}{Finesse $F$}   & $2\pi/\left(\mathcal{T}_1+\mathcal{T}_2+\mathcal{L}\right)$  & $1.4\times 10^4$ &  $2.5\times 10^3$ \\ 
    \multicolumn{1}{c}{$\kappa_{ex}$}   & $c \mathcal{T}_1/(4\ell)$     & 2$\pi\times$ 179 kHz &   2$\pi\times$ 45 MHz   \\    \multicolumn{1}{c}{$\kappa_i$}     & $c \left(\mathcal{T}_2+\mathcal{L}\right)/(4\ell)$ & 2$\pi\times$ 90 kHz  &     2$\pi\times$ 30 MHz \\ 
    \hline 
    \multicolumn{1}{c}{Cooperativity $C_t$}     & $g^2/(\kappa_t\gamma)$ & 3.1  &     2.2  \\
    \hline 
    \multirow{2}{2.2cm}{Estimated gate time operation} \vspace{1mm}     & $\simeq 3\max\{1/\kappa_t,1/C\gamma\}$ \vspace{1mm} & $2~\mu$s \vspace{1mm} &     20 ns \vspace{1mm} \\
    \hline
    \hline
    \end{tabular}
\end{center}
\caption{\label{tab:table2} Possible cavity parameters for the two above configurations of Table~\ref{tab:table}, at $\lambda = 370$~nm, with $c$ the speed of light. }
\end{table*}

Fig.~\ref{fig:Fig_Sym} shows the optimized parameters $\delta_c^{\rm{opt}}$ (a) and $\delta_a^{\rm{opt}}$ (b) as a function of $\kappa_{ex}$ characterizing the coupling mirror transmission, in the range where the total cooperativity $C_{t}=\cfrac{\left(\chi g\right)^2}{\kappa_{t}\gamma}$ is bigger than one. 
In accordance with Eq.~(\ref{solutions2}), two sets of solutions are shown, in solid and dashed lines. 
We introduce the three following coupling rates: $\kappa_{ex}^{\rm{(I)}}$ associated to a maximal $\delta_c$ (labelled as $\delta_c^{\rm{(I)}}$), $\kappa_{ex}^{\rm{(II)}} = \kappa_{ex}^{\rm{opt}}$, and $\kappa_{ex}^{\rm{(III)}}$ associated to $C_t =1$. 

The associated optimized fidelity and efficiency are shown in solid blue lines in the frames (c) and (d); they reduce to the black dashed lines if no optimization is performed, i.e. $\delta_c=\delta_a=0$. 
In accordance with the comments of Sec.~\ref{sec:model}, the fidelity reaches unity by tuning $\delta_c$ and $\delta_a$ to $\delta_c^{\mathrm{opt}}$ and $\delta_a^{\mathrm{opt}}$ when $\kappa_i \leq \kappa_{ex} \leq \kappa_{ex}^{\rm{opt}}$ (white area in Fig.~\ref{fig:Fig_Sym}). 
In this coupling range and for $C_t \gg 1$, an analytic derivation shows that the fidelity increases by a quantity that equals $\Delta \mathcal{F} = \cfrac{A}{1+A}$ with $A =P_D/P_B \simeq \abs{\cfrac{\kappa_i}{\kappa_{ex}} - \cfrac{1}{2C_t}}^2$ compared to the case where no detuning is applied.

Fig.~\ref{fig:Fig_Sym} can be directly used for any cavity and ionic transition parameters by replacing the various variables with the analytical expressions reported in Table~\ref{tab:table}. As an example, we give the numerical values of these variables in the case of the ionic transition considered here and of a typical macroscopic Fabry-Perot resonator ($\kappa_i = 2\pi\times 90$~kHz and $\kappa_i \leq \kappa_{ex}\leq 2\pi\times 743$~kHz) with an achievable coherent coupling rate $g = 2\pi\times 5$~MHz \cite{Stute2012,Stute2013}, and in the case of a realistic fiber-based Fabry-Perot resonator ($\kappa_i = 2\pi\times 30$~MHz and $\kappa_i \leq \kappa_{ex} \leq 2\pi\times 133$~MHz) with a coherent coupling rate $g = 2\pi\times 70$~MHz, high but already attained \cite{Ballance2017}. 
The cavity parameters needed to achieve the experimental situations of Table~\ref{tab:table} are given in Table~\ref{tab:table2}: specifically the length $\ell$ of the Fabry-Perot resonator, its input-output coupling mirror transmission $\mathcal{T}_1$, back mirror transmission $\mathcal{T}_2$ and other intrinsic losses such as mirrors absorption and scattering $\mathcal{L}$. 

\begin{figure}%
\begin{center}
\includegraphics[scale=0.4]{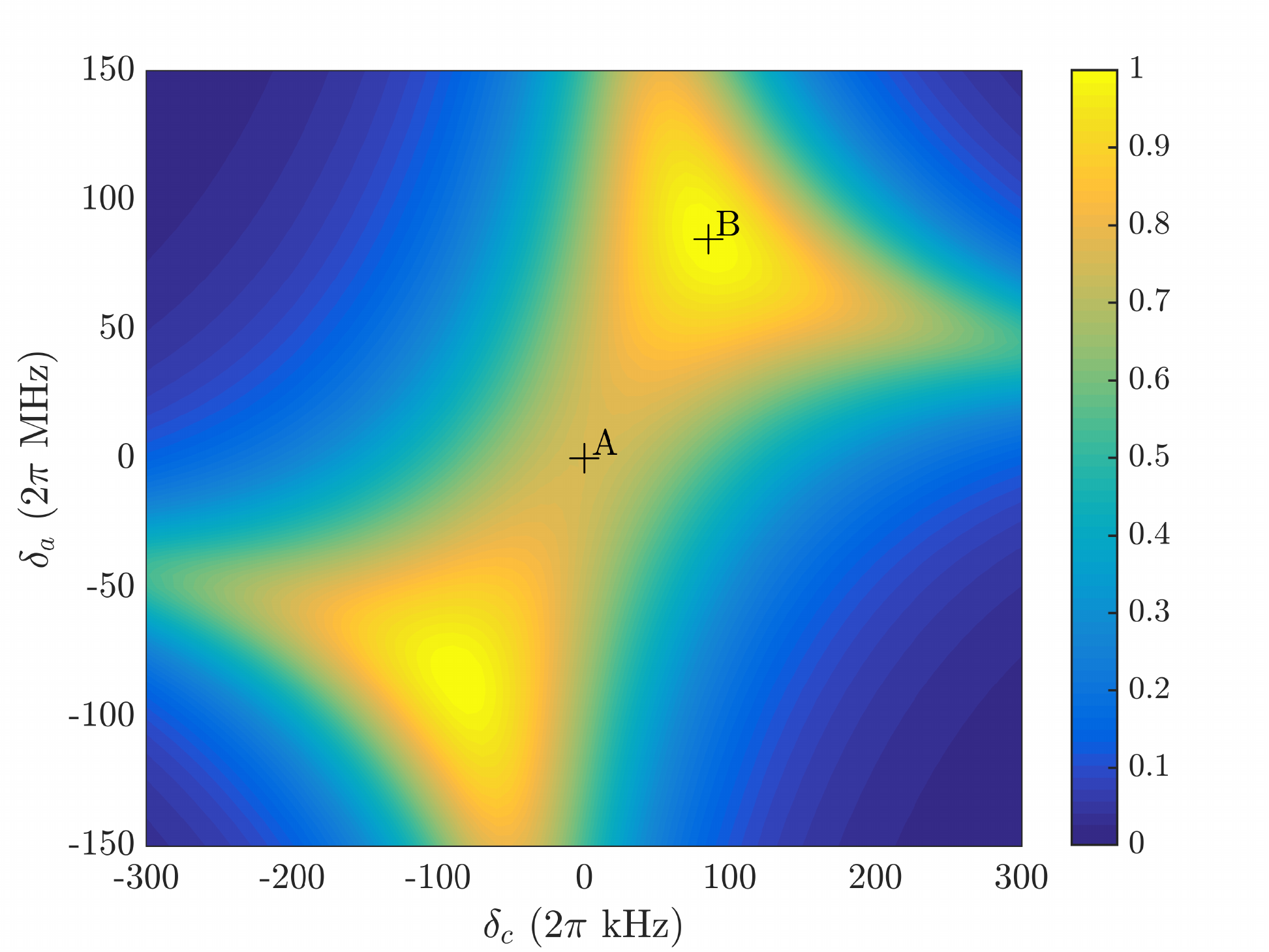} 
\end{center}
\vspace{-4mm}
\caption{Map of the SWAP fidelity as a function of the cavity and atomic detunings $\delta_c$ and $\delta_a$ for the $^2S_{1/2} (F=1)$ - $^2P_{1/2} (F'=0)$ transition of \textsuperscript{171}Yb\textsuperscript{+} and extrinsic coupling $\kappa_{ex}=2\pi\times 135$  kHz: the point A indicates $\delta_c=\delta_a=0$ and leads to $\mathcal{F}\simeq 0.75$, and the point B indicates $\delta_c^{\rm{opt}}$ and $\delta_a^{\rm{opt}}$ and leads to $\mathcal{F}=1$. }
\label{fig:Fig_Sym_map}%
\end{figure}

Fig.~\ref{fig:Fig_Sym_map} shows a map of the fidelity as a function of the parameters $\delta_c$ and $\delta_a$ for a given $\kappa_{ex}=2\pi\times 135$ kHz: the optimization (point B) leads to a fidelity equal to 1, i.e. an increase of 25$\%$ compared to the situation where no detuning is applied (point A). 
\\

Finally, note that the gate operation time, here set by the seeder pulse duration $1/\kappa_s$, needs to be significantly longer than both the cavity lifetime $1/\kappa_t$ and the enhanced atomic response time $1/(C\gamma)$. 
In the above example of the conventional Fabry-Perot resonator (Table~\ref{tab:table2}), $\tau_{\rm{SWAP}}$ is bound from below by the cavity decay time ($\simeq 600$~ns), i.e. of the order of a few $\mu$s.  In the case of the fiber-based Fabry-Perot resonator, $\tau_{\rm{SWAP}}$ is bound from below by the cavity-enhanced spontaneous emission time ($\simeq 7$~ns), i.e. of the order of a few tens of ns.  
In both cases, the length of the cavity affects the operation time of the gate: either through the cavity lifetime, $\cfrac{1}{\kappa_t} = \cfrac{4l}{c(\mathcal{T}_1+\mathcal{T}_2+\mathcal{L})}$, or through the cavity-enhanced decay time, $\cfrac{1}{C\gamma} = \cfrac{c}{4lg^2}\left(\cfrac{1}{\mathcal{T}_1}+\cfrac{1}{\mathcal{T}_2+\mathcal{L}}\right)^{-1}$.   
\\

\subsection{Asymmetric $\Lambda$ system}
\label{subsec:3-2}

\begin{figure}[h]%
\centering
\subfigure{%
\label{fig:Ca_levels_all}%
\includegraphics[scale=0.33]{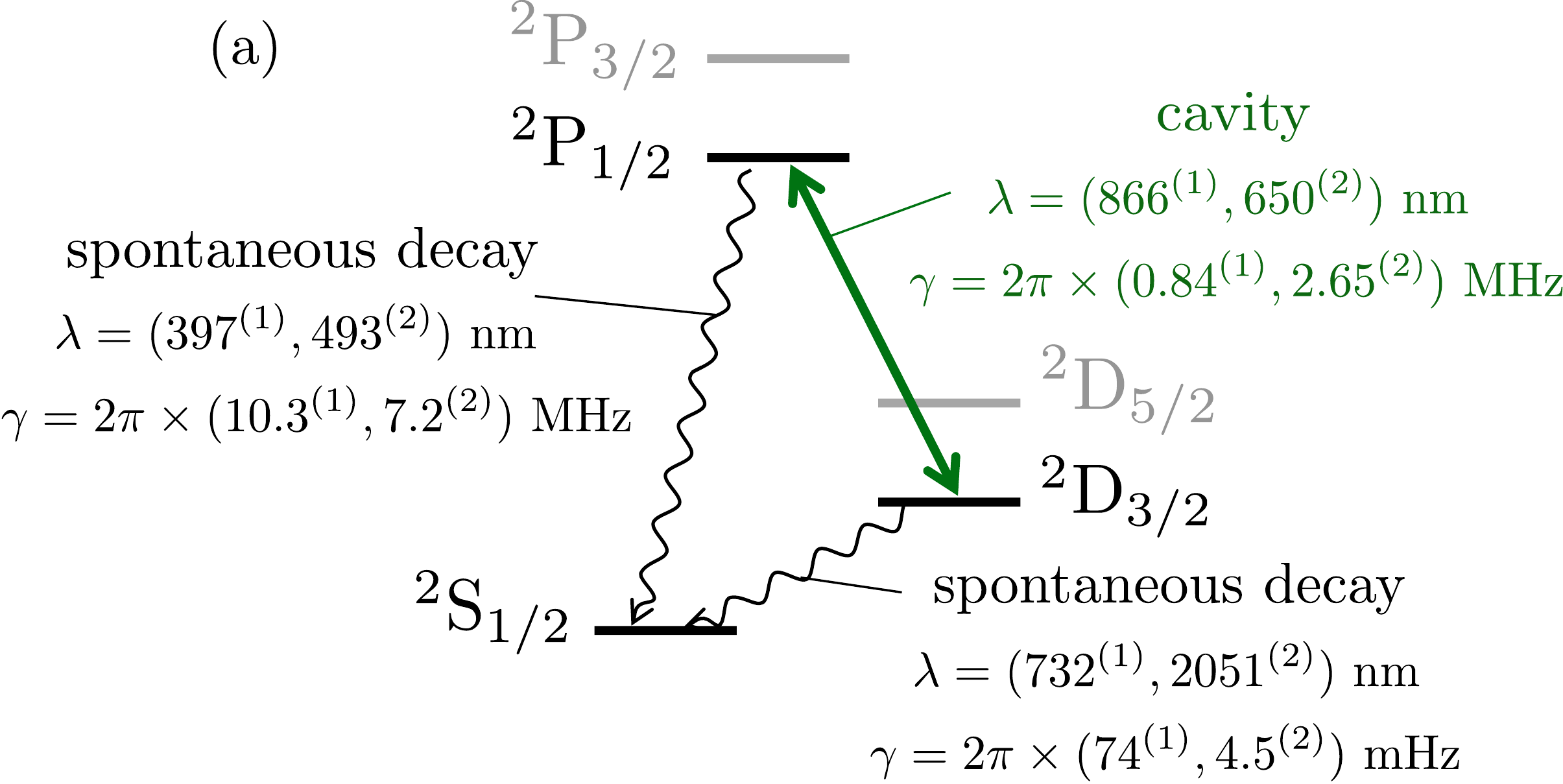}}%
\qquad
\subfigure{%
\label{fig:Ca_levels_lambda}%
\includegraphics[scale=0.33]{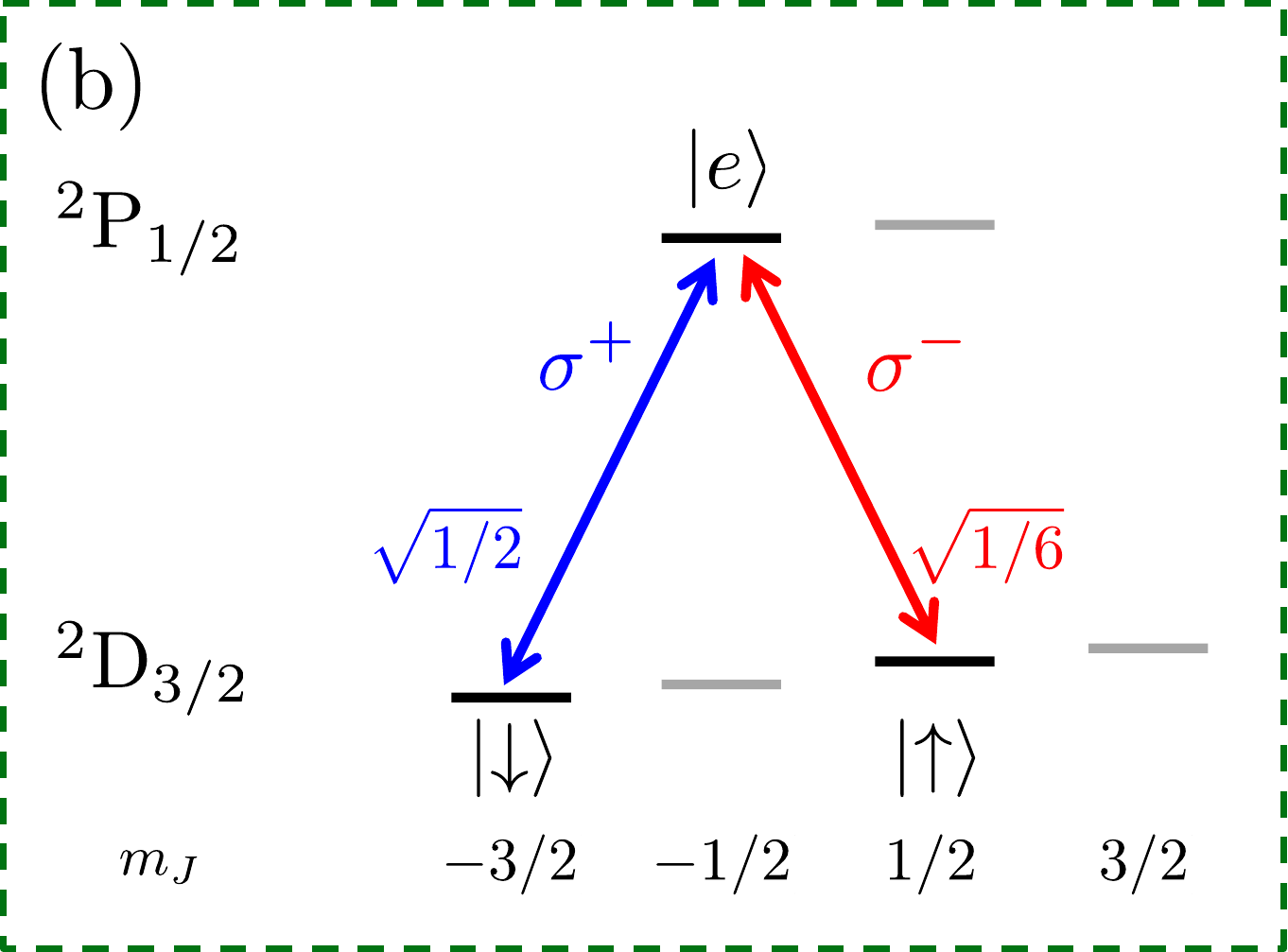}}%
\qquad
\subfigure{%
\label{fig:Ca_Zeeman}%
\includegraphics[scale=0.2]{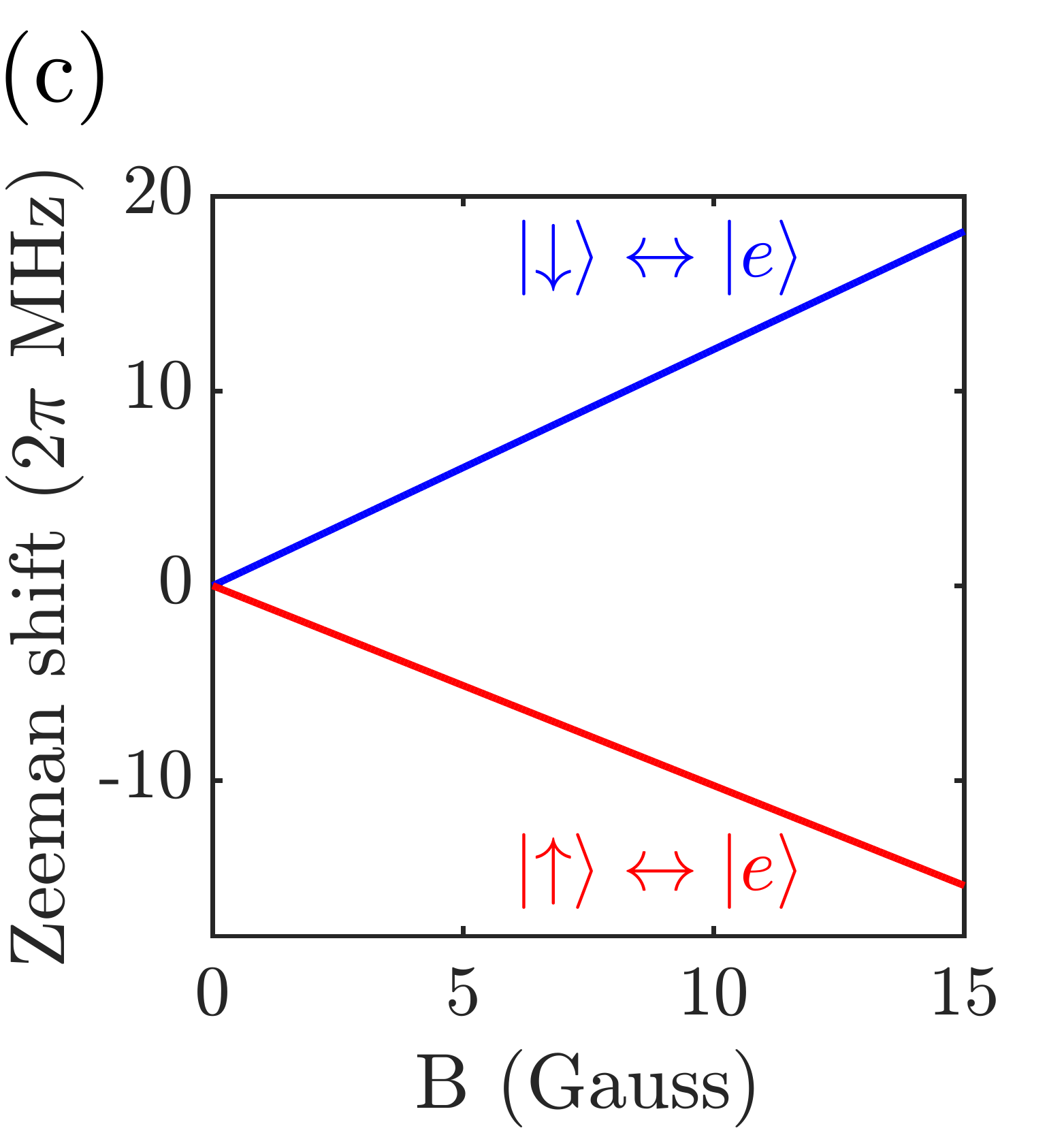}}%
\vspace{-2mm}
\caption{Configuration for a SWAP gate with an asymmetric $\Lambda$ system (unequal transition strengths). (a) Energy levels, wavelengths  and decay rates of  \textsuperscript{40}Ca\textsuperscript{+} (superscript (1)) and \textsuperscript{138}Ba\textsuperscript{+} (superscript (2)). 
(b) Three-level $\Lambda$ system considered here involving the \textsuperscript{2}D\textsubscript{3/2} and \textsuperscript{2}P\textsubscript{1/2} manifolds, and (c) Zeeman splitting of the two relevant transitions as a function of the magnetic field $B$. }
\label{fig:Ca_levels_main}
\end{figure}
Next, we consider the \textsuperscript{2}D\textsubscript{3/2} - \textsuperscript{2}P\textsubscript{1/2} transition of a \textsuperscript{40}Ca\textsuperscript{+} ion at 866 nm or \textsuperscript{138}Ba\textsuperscript{+} at 650 nm, where $\left(\gamma_{\rm{Ca}}, \gamma_{\rm{Ba}}\right) = 2\pi\times \left(11.1, 9.9 \right)$~MHz $\simeq 2\pi\times 10$ MHz (Fig.~\ref{fig:Ca_levels_all}). At such wavelengths, the cavity losses are much smaller than in the previous ultraviolet case, which allows to consider smaller mirror transmissions coefficients and smaller, state-of-the-art $g$ parameters (see Table~\ref{tab:table3}). 
As pictured in Fig.~\ref{fig:Ca_levels_lambda}, a $\Lambda$ system can be isolated from the level structure by considering for instance the two transitions $\ket{\downarrow=\{m_J=-3/2\}}-\ket{e=\{m_J'=-1/2\}}$ and $\ket{\uparrow=\{m_J=1/2\}}-\ket{e=\{m_J'=-1/2\}}$ via an initial classical preparation, with no possible leakage to the other Zeeman sublevels. Note however that the \textsuperscript{2}D\textsubscript{3/2} manifold being metastable ($\tau_{\rm{Ca}} = 1$~sec, $\tau_{\rm{Ba}} = 18$~sec), the ion can decay to the $^2S_{1/2}$ manifold. Nonetheless, discarding this event by postselection will not affect the fidelity of SPRINT. Here the Clebsch-Gordan coefficients are $\left(\chi_{\downarrow},\chi_{\uparrow}\right)=(\sqrt{1/2},\sqrt{1/6})$: in this asymmetric configuration, the optimization of the fidelity can benefit from the use of an external magnetic field. The Zeeman shifts associated with these transitions are shown in Fig.~\ref{fig:Ca_Zeeman}. 
\begin{table*}[h!]
\begin{center}
\renewcommand{\arraystretch}{1.5}
    \begin{tabular}{| m{4cm}  | >{\centering\arraybackslash}m{3.8cm}   >{\centering\arraybackslash}m{3.8cm}    }
    \hline
	\hline
    \multicolumn{1}{c}{}    & typical conventional Fabry-Perot cavity    & realistic fiber-based Fabry-Perot cavity  \\ 
	\hline 
    \multicolumn{1}{c}{$\ell$}                & 20 mm & 400 $\mu$m \\ 
    \multicolumn{1}{c}{$\mathcal{T}_1$}    & 50 ppm & 600 ppm \\ 
    \multicolumn{1}{c}{$\mathcal{T}_2 + \mathcal{L}$}    & 17 ppm & 100 ppm \\ 
    \hline
    \multicolumn{1}{c}{Finesse $F$}    & $9.4\times 10^4$ &  $9.0\times 10^3$ \\ 
    \multicolumn{1}{c}{$\kappa_{ex}$}   & 2$\pi\times$ 30 kHz &   2$\pi\times$ 18 MHz   \\ 
    \multicolumn{1}{c}{$\kappa_i$}    & 2$\pi\times$ 10 kHz  &     2$\pi\times$ 3 MHz  \\
    \hline 
    \multicolumn{1}{c}{$\left(g_{\downarrow}, g_{\uparrow}\right)=\left(\chi_\downarrow,\chi_\uparrow\right) \times g$}   & 2$\pi\times \left(1.4, 0.82\right)$ MHz &  2$\pi\times \left(28, 16\right)$ MHz \\ 
    \multicolumn{1}{c}{$\gamma$}    & 2$\pi\times$ 10 MHz & 2$\pi\times$ 10 MHz \\ 
     \multicolumn{1}{c}{Cooperativity $C$}      & 3.3  &     2.4  \\
         \hline
     \multicolumn{1}{c}{Estimated gate operation time}     & 2 $\mu$s &     20 ns \\

    \hline
    \hline
    \end{tabular}
\end{center}
\caption{\label{tab:table3} Possible parameters for conventional and fiber-based Fabry-Perot cavities with a resonance frequency in the visible or near infrared. The considered atomic transition is the \textsuperscript{2}D\textsubscript{3/2} to \textsuperscript{2}P\textsubscript{1/2} of a \textsuperscript{40}Ca\textsuperscript{+} or \textsuperscript{138}Ba\textsuperscript{+} ion. }
\end{table*}
\\

\begin{figure}[h!]%
\begin{center}
\label{fig:Fig_Asym_opt_2}%
\includegraphics[scale=0.35]{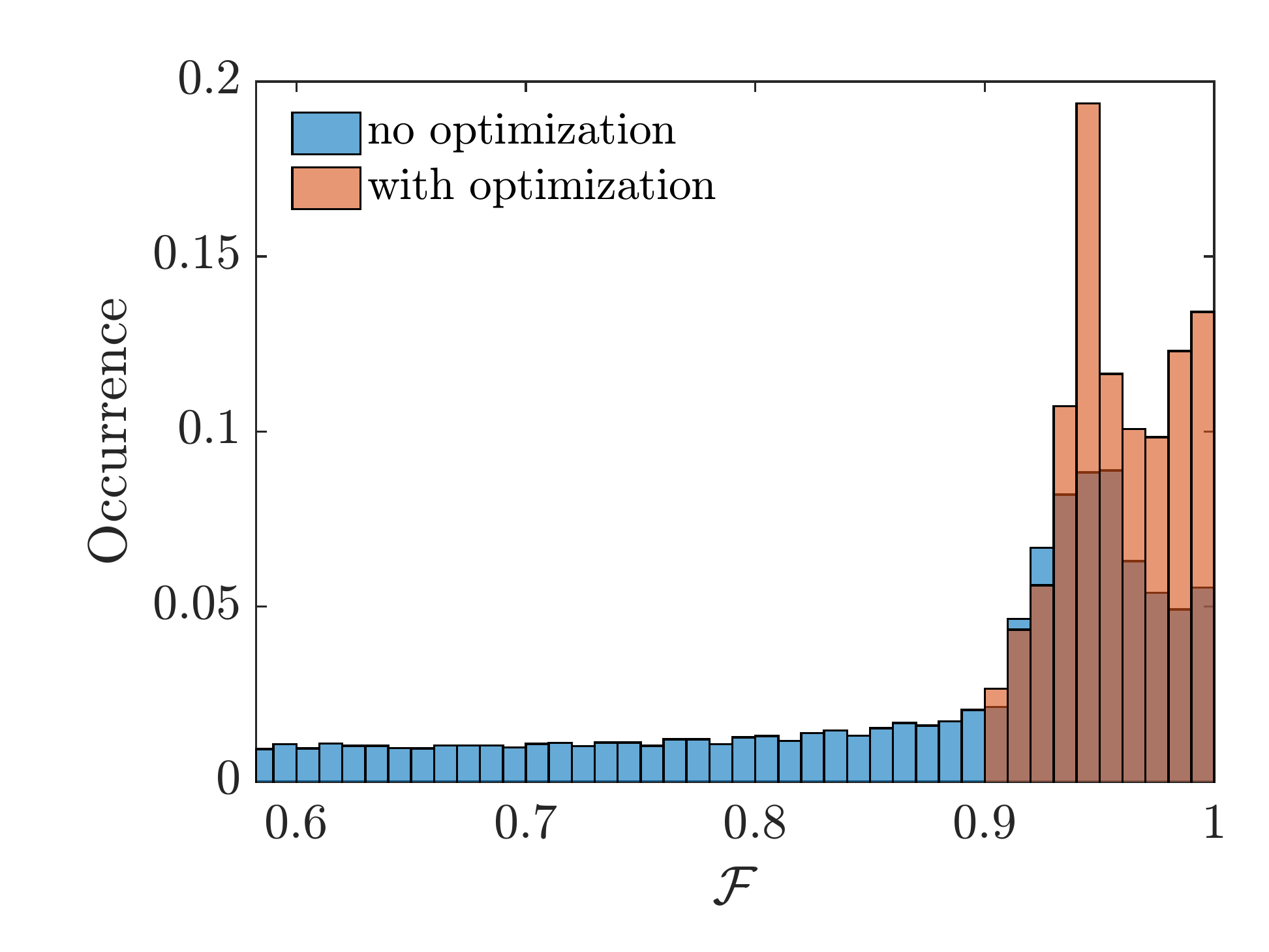}%
\end{center}
\vspace*{-7mm}
\caption{Fidelity distributions when the initial photonic and atomic qubit parameters $\left(\alpha,\alpha'\right)$ and $\left(\beta,\beta'\right)$ span the Bloch spheres. The cavity parameters are given in Table~\ref{tab:table3}. Blue histogram: no optimization; red histogram: with numerically optimized cavity detuning $\delta_c^{\rm{opt}}=2\pi\times 18$ kHz, atomic detuning $\delta_a^{\rm{opt}}=2\pi\times 46$ MHz and magnetic field $B^{\rm{opt}}=8.5$ G. }
\label{fig:Fig_Asym_opt}%
\end{figure}

Following Sec.~\ref{sec:model}, a numerical optimization of the parameters $\delta_c$, $\delta_a$ and $B$ is performed in order to maximize the  fidelity of the SWAP gate for arbitrary input optical and material qubits. 
The distribution of fidelities arising from these different initial qubits is displayed in Fig.~\ref{fig:Fig_Asym_opt} in the case of a macroscopic Fabry-Perot cavity (cf. Table~\ref{tab:table3}). 
With the optimized parameters $\delta_c^{\rm{opt}}=2\pi\times 18$ kHz, $\delta_a^{\rm{opt}}=2\pi\times 46$ MHz and $B^{\rm{opt}}=8.5$ G, the distribution (red histogram) shows an average of $\overline{\mathcal{F}}^{\rm{opt}} = 95.8 \pm 2.5 \%$, to be compared to $\overline{\mathcal{F}}^{0} = 87.7 \pm 11.3 \%$ in the case where no optimization is performed (blue histogram). 
\begin{figure*}[h!]%
\begin{center}
\hspace{-4mm}
\includegraphics[scale=0.48]{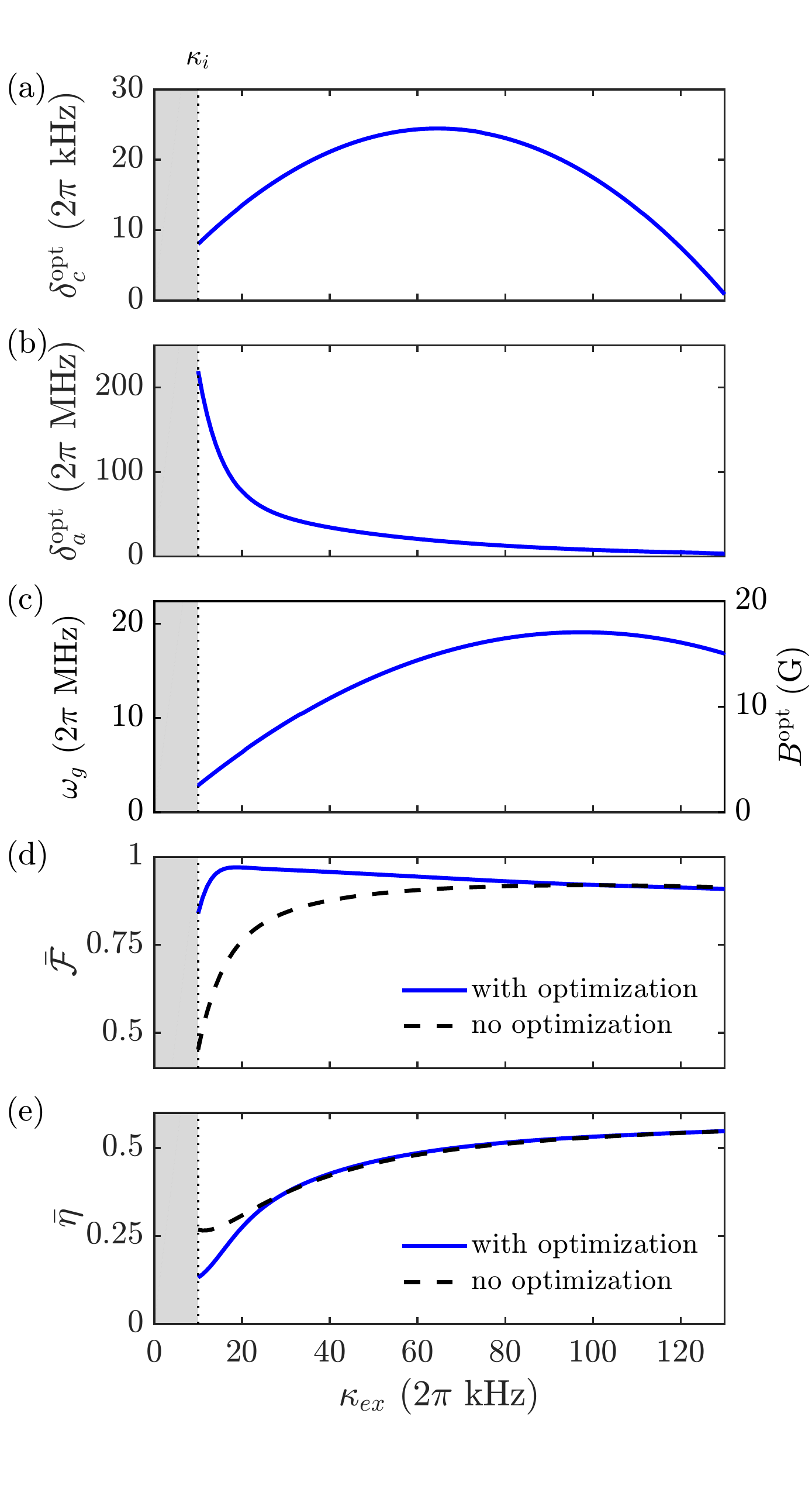} 
\includegraphics[scale=0.48]{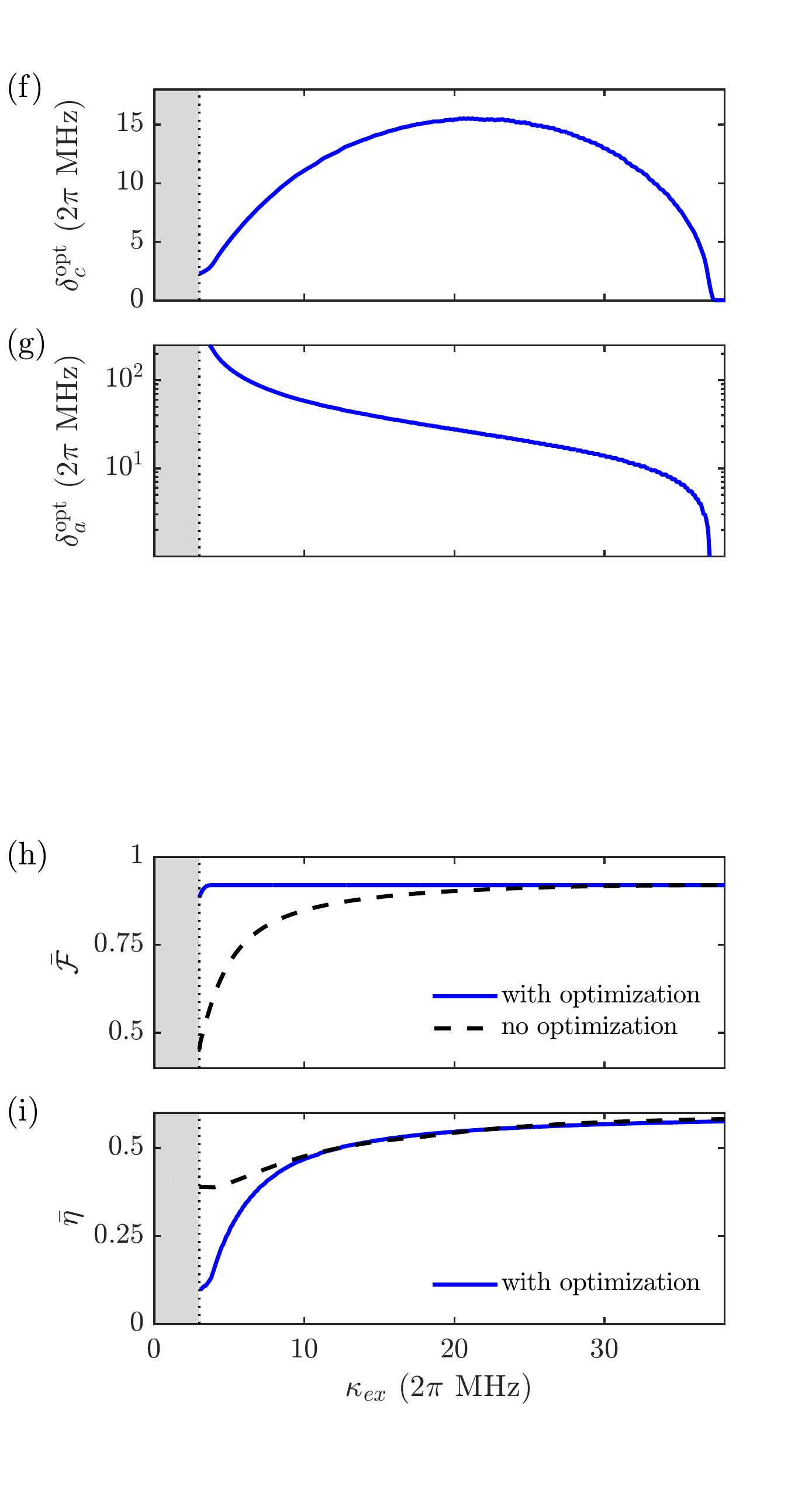} 
\end{center}
\vspace*{-10mm}
\caption{Optimal cavity detunings $\delta_c^{\rm{opt}}$ (frames a and f), atomic detunings $\delta_a^{\rm{opt}}$ (b and g), Larmor frequencies $\omega_g^{\rm{opt}}$ and corresponding $B^{\rm{opt}}$ (c) as a function of the extrinsinc coupling $\kappa_{ex}$. The resulting average fidelities $\overline{\mathcal{F}}$ (d and h) and average efficiencies $\overline{\eta}$ (e and i) are shown as solid blue lines, while the non-optimized fidelities and efficiencies are shown as dashed black lines. On the left (frames (a) to (e)), typical conventional Fabry-Perot; on the right (frames (f) to (i)),  $B=0$ G and realistic fiber-based Fabry-Perot. The system parameters are given in Table~\ref{tab:table3}. }
\label{fig:Fig_Asym_kex}%
\end{figure*}

In Fig.~\ref{fig:Fig_Asym_kex}, we consider two settings examples corresponding to a macroscopic cavity (frames (a) to (e)) and a fiber-based cavity (frames (f) to (i)), and perform the numerical optimization of the fidelity averaged over the input qubits as a function of $\kappa_{ex}$. We display the optimal parameters $\delta_c^{\rm{opt}}$, $\delta_a^{\rm{opt}}$ and $B^{\rm{opt}}$ (or equivalently the Larmor frequency $\omega_g^{\rm{opt}}$), the optimized and non-optimized fidelities, and corresponding efficiencies. 
In the conventional Fabry-Perot configuration, the maximal increase in the average fidelity is found at $\kappa_{ex} = 2\pi \times 19$~kHz and amounts to $22~\%$ (from $\overline{\mathcal{F}}^{0}=75\%$ to $\overline{\mathcal{F}}^{\rm{opt}}=97\%$), while the associated efficiency decreases by $4~\%$ (from $\overline{\eta}^0=30\%$ to $\overline{\eta}^{\rm{opt}}=26\%$). 
In the fiber-based Fabry-Perot configuration, the optimal applied magnetic fields would be too high to be experimentally reasonable (typically up to several kG): this originates from the higher intrinsic losses of such cavities. We then chose to perform the optimization with no applied magnetic field, which still leads in this case to a fair improvement of the fidelity to a mostly identical value of $\overline{\mathcal{F}}^{\rm{opt}} = 92~\%$. For instance, choosing $\kappa_{ex}=2\pi\times 8$~MHz leads to an increase in fidelity of $11~\%$ and a decrease in efficiency of $3~\%$. 

As in the previous section, the gate operation time in the case of a conventional Fabry-Perot resonator is here again bounded from below by the cavity decay time and equals a few $\mu$s. In the case of a fiber-based Fabry-Perot resonator, $\kappa_t^{-1} \simeq \left(C_t \gamma\right)^{-1}$ and $\tau_{\rm{SWAP}}$ is of the order a few tens of ns.

\subsection{Deviation from an isolated $\Lambda$ system}
\label{subsec:4.3}

In this last section, we comment on the fact that a simple, ideal $\Lambda$-type three-level atom coupled solely to two optical modes is unrealistic. The three levels of the $\Lambda$-systems considered in this work are Zeeman sublevels that have been chosen such that selection rules should prevent the ion to decay to unwanted sub-levels. However, a photon with an imperfect polarization state could drive a "wrong" transition. An exhaustive study of such imperfections has been performed in Ref. \cite{Rosenblum2017}. Note that the Fabry-Perot setup considered here should be more robust to these imperfections (such as polarization mismatch, cross-talk between optical modes, presence of near-by excited states) than that considered in Ref. \cite{Rosenblum2017}. 

Additionally, the ion may decay to a different hyperfine manifold after being excited. These events have to be discarded in order to keep the fidelity of the process unharmed, at the price of lowering the efficiency. To do so, each realization of the proposed scheme must therefore include post-selection (e.g. through spectral filtering) and initial atomic repumping to the ground state manifold.  
The efficiency of the photon-ion process $\eta$ considered previously is then modified as follows: 
\begin{align}
\label{eta_tilde}
\tilde{\eta} = \eta \times \cfrac{\Gamma}{\Gamma + \gamma_\mathrm{other}}
\end{align}
where $\Gamma = C\gamma$ is the cavity-enhanced spontaneous emission rate for the hyperfine transition of interest and $\gamma_\mathrm{other}$ is the spontaneous emission rate corresponding to all other transitions. As an illustration, we take the example provided in Sec. \ref{sec:ions}\ref{subsec:3-2} for \textsuperscript{138}Ba\textsuperscript{+} (\textsuperscript{40}Ca\textsuperscript{+}) and a cooperativity $C=3$, compatible with Table \ref{tab:table3}: with $\Gamma/(2\pi) = 8.0$ ($2.5$) MHz corresponding to the decay rate of the hyperfine transition \textsuperscript{2}D\textsubscript{3/2} - \textsuperscript{2}P\textsubscript{1/2} and $\gamma_\mathrm{other}/(2\pi) = 7.2$ ($10.3$) MHz to \textsuperscript{2}S\textsubscript{1/2} - \textsuperscript{2}P\textsubscript{1/2} , we find $\tilde{\eta} = 0.52 \eta$ ($0.20 \eta$). 

Finally, the ground states of the three-level atom may be metastable (as it is the case in Sec. \ref{sec:ions}\ref{subsec:3-2}), allowing a decay channel to the ground state. Generally, this supports the need for an initial atomic preparation, although in practice the associated relaxation rates are about 7 to 9 orders of magnitude smaller than $\Gamma$, hence than the gate timescale, for the ions considered in this work. These processes therefore have a negligible contribution on the modification of the efficiency. 

\section{Summary}
\label{sec:ccl}

This paper demonstrated the feasibility of implementing an ion-photon qubit SWAP gate in realistic trapped ion systems, based on the deterministic 
single-photon Raman interaction. Importantly, this scheme requires Purcell enhancement but not necessarily strong coupling: in other words, it enables the use of cavities with small mode area yet a reasonably long distance between the mirrors, which is favorable so that the ion trap potential remains undisturbed. 

This theoretical analysis gave the framework to the swap protocol, in particular by discussing the relevant parameters leading to optimize its performance. Specifically, in the case of an equally weighted three-level $\Lambda$ system, we showed that there exists a range of extrinsic coupling rates where an appropriate tuning of the probe-cavity and probe-atom frequency detunings restores the fidelity to unity, however at the price of a decrease in the efficiency. In addition to these detunings, the degeneracy of the Zeeman sublevels can be lifted to further optimize the fidelity in the case of an unequally weighted three-level $\Lambda$ system. We quantitatively applied our model to realistic systems, involving \textsuperscript{171}Yb\textsuperscript{+}, \textsuperscript{40}Ca\textsuperscript{+} and \textsuperscript{138}Ba\textsuperscript{+} ions. We showed that the implementation of a SPRINT-based SWAP gate is realistic in both state-of-the-art conventional and fiber-based Fabry-Perot cavities.
This scheme, highly scalable and ns-fast, is therefore a powerful building block that can be further exploited to realize photon-photon quantum gates, in particular universal ones such as $\sqrt{\rm{SWAP}}$ \cite{Koshino2010} or C-phase \cite{Tokunaga2015}.

\section*{Acknowledgements}
\label{sec:ack}
A. B. and B. D. acknowledge support from the Israeli Science Foundation, the Minerva Foundation and the Crown Photonics Center. This research was made possible in part by the historic generosity of the Harold Perlman family.
T. N. and R. B. acknowledge financial support by the U.S. Army Research Laboratory’s Center for Distributed Quantum Information via the project SciNet: Scalable Ion-Trap Quantum Network, Cooperative Agreement No. W911NF15-2-0060, the Austrian Science Fund (FWF) through Project F 7109, and the European Union's Horizon 2020 research and innovation program under grant agreement No. 820445 and project name 'Quantum Internet Alliance'. 

\section*{Disclosures}
The authors declare no conflicts of interest.

\bibliography{mybib_ions}

\begin{thebibliography}{10}
\newcommand{\enquote}[1]{``#1''}

\bibitem{Kimble2008}
H.~J. Kimble, \enquote{The quantum internet,} {\protect\JournalTitle{Nature}}
  \textbf{453}, 1023--1030 (2008).

\bibitem{Luo2009}
L.~Luo, D.~Hayes, T.~Manning, D.~Matsukevich, P.~Maunz, S.~Olmschenk, J.~Sterk,
  and C.~Monroe, \enquote{Protocols and techniques for a scalable atom--photon
  quantum network,} {\protect\JournalTitle{Fortschritte der Physik}}
  \textbf{57}, 1133--1152 (2009).

\bibitem{Duan2010}
L.-M. Duan and C.~Monroe, \enquote{Colloquium: {Q}uantum networks with trapped
  ions,} {\protect\JournalTitle{Rev. Mod. Phys.}} \textbf{82}, 1209--1223
  (2010).

\bibitem{Monroe2013}
C.~Monroe and J.~Kim, \enquote{Scaling the ion trap quantum processor,}
  {\protect\JournalTitle{Science}} \textbf{339}, 1164--1169 (2013).

\bibitem{Northup2014}
T.~Northup and R.~Blatt, \enquote{Quantum information transfer using photons,}
  {\protect\JournalTitle{Nat. Photon.}} \textbf{8}, 356--363 (2014).

\bibitem{Barrett2004}
M.~D. Barrett, J.~Chiaverini, T.~Schaetz, J.~Britton, W.~M. Itano, J.~D. Jost,
  E.~Knill, C.~Langer, D.~Leibfried, R.~Ozeri, and D.~J. Wineland,
  \enquote{Deterministic quantum teleportation of atomic qubits,}
  {\protect\JournalTitle{Nature}} \textbf{429}, 737--739 (2004).

\bibitem{Riebe2004}
M.~Riebe, H.~H{\"a}ffner, C.~F. Roos, W.~H{\"a}nsel, J.~Benhelm, G.~P.~T.
  Lancaster, T.~W. K{\"o}rber, C.~Becher, F.~Schmidt-Kaler, D.~F.~V. James, and
  R.~Blatt, \enquote{Deterministic quantum teleportation with atoms,}
  {\protect\JournalTitle{Nature}} \textbf{429}, 734--737 (2004).

\bibitem{Olmschenk2009}
S.~Olmschenk, D.~Matsukevich, P.~Maunz, D.~Hayes, L.-M. Duan, and C.~Monroe,
  \enquote{Quantum teleportation between distant matter qubits,}
  {\protect\JournalTitle{Science}} \textbf{323}, 486--489 (2009).

\bibitem{kurz2014}
C.~Kurz, M.~Schug, P.~Eich, J.~Huwer, P.~M{\"u}ller, and J.~Eschner,
  \enquote{Experimental protocol for high-fidelity heralded photon-to-atom
  quantum state transfer,} {\protect\JournalTitle{Nature communications}}
  \textbf{5}, 1--5 (2014).

\bibitem{Pirandola2015}
S.~Pirandola, J.~Eisert, C.~Weedbrook, A.~Furusawa, and S.~L. Braunstein,
  \enquote{Advances in quantum teleportation,} {\protect\JournalTitle{Nature
  photonics}} \textbf{9}, 641--652 (2015).

\bibitem{kurz2016}
C.~Kurz, P.~Eich, M.~Schug, P.~M{\"u}ller, and J.~Eschner,
  \enquote{Programmable atom-photon quantum interface,}
  {\protect\JournalTitle{Physical Review A}} \textbf{93}, 062348 (2016).

\bibitem{Wan2019}
Y.~Wan, D.~Kienzler, S.~D. Erickson, K.~H. Mayer, T.~R. Tan, J.~J. Wu, H.~M.
  Vasconcelos, S.~Glancy, E.~Knill, D.~J. Wineland, A.~C. Wilson, and
  D.~Leibfried, \enquote{Quantum gate teleportation between separated qubits in
  a trapped-ion processor,} {\protect\JournalTitle{Science}} \textbf{364},
  875--878 (2019).

\bibitem{Herskind2009}
P.~F. Herskind, A.~Dantan, J.~P. Marler, M.~Albert, and M.~Drewsen,
  \enquote{Realization of collective strong coupling with ion {C}oulomb
  crystals in an optical cavity,} {\protect\JournalTitle{Nat. Phys.}}
  \textbf{5}, 494--498 (2009).

\bibitem{Hunger2010}
D.~Hunger, T.~Steinmetz, Y.~Colombe, C.~Deutsch, T.~W. H{\"a}nsch, and
  J.~Reichel, \enquote{A fiber {F}abry--{P}erot cavity with high finesse,}
  {\protect\JournalTitle{New J. Phys.}} \textbf{12}, 065038 (2010).

\bibitem{Stute2012}
A.~Stute, B.~Casabone, B.~Brandst{\"a}tter, D.~Habicher, H.~Barros, P.~Schmidt,
  T.~Northup, and R.~Blatt, \enquote{Toward an ion--photon quantum interface in
  an optical cavity,} {\protect\JournalTitle{Appl. Phys. B}} \textbf{107},
  1145--1157 (2012).

\bibitem{Sterk2012}
J.~Sterk, L.~Luo, T.~Manning, P.~Maunz, and C.~Monroe, \enquote{Photon
  collection from a trapped ion-cavity system,} {\protect\JournalTitle{Phys.
  Rev. A}} \textbf{85}, 062308 (2012).

\bibitem{Brandstatter2013}
B.~Brandst{\"a}tter, A.~McClung, K.~Sch{\"u}ppert, B.~Casabone, K.~Friebe,
  A.~Stute, P.~O. Schmidt, C.~Deutsch, J.~Reichel, R.~Blatt, and T.~E. Northup,
  \enquote{Integrated fiber-mirror ion trap for strong ion-cavity coupling,}
  {\protect\JournalTitle{Rev. Sci. Instr.}} \textbf{84}, 123104 (2013).

\bibitem{Steiner2013}
M.~Steiner, H.~M. Meyer, C.~Deutsch, J.~Reichel, and M.~K{\"o}hl,
  \enquote{Single ion coupled to an optical fiber cavity,}
  {\protect\JournalTitle{Phys. Rev. Lett.}} \textbf{110}, 043003 (2013).

\bibitem{Stute2013}
A.~Stute, B.~Casabone, B.~Brandst{\"a}tter, K.~Friebe, T.~Northup, and
  R.~Blatt, \enquote{Quantum-state transfer from an ion to a photon,}
  {\protect\JournalTitle{Nat. Photon.}} \textbf{7}, 219--222 (2013).

\bibitem{Takahashi2013}
H.~Takahashi, A.~Wilson, A.~Riley-Watson, F.~Oru{\v{c}}evi{\'c},
  N.~Seymour-Smith, M.~Keller, and W.~Lange, \enquote{An integrated fiber trap
  for single-ion photonics,} {\protect\JournalTitle{New J. Phys.}} \textbf{15},
  053011 (2013).

\bibitem{Cetina2013}
M.~Cetina, A.~Bylinskii, L.~Karpa, D.~Gangloff, K.~M. Beck, Y.~Ge, M.~Scholz,
  A.~T. Grier, I.~Chuang, and V.~Vuleti{\'c}, \enquote{One-dimensional array of
  ion chains coupled to an optical cavity,} {\protect\JournalTitle{New J.
  Phys.}} \textbf{15}, 053001 (2013).

\bibitem{Steiner2014}
M.~Steiner, H.~Meyer, J.~Reichel, and M.~K{\"o}hl, \enquote{Photon emission and
  absorption of a single ion coupled to an optical-fiber cavity,}
  {\protect\JournalTitle{Phys. Rev. Lett.}} \textbf{113}, 263003 (2014).

\bibitem{Casabone2015}
B.~Casabone, K.~Friebe, B.~Brandst{\"a}tter, K.~Sch{\"u}ppert, R.~Blatt, and
  T.~Northup, \enquote{Enhanced quantum interface with collective ion-cavity
  coupling,} {\protect\JournalTitle{Phys. Rev. Lett.}} \textbf{114}, 023602
  (2015).

\bibitem{Eltony2016}
A.~M. Eltony, D.~Gangloff, M.~Shi, A.~Bylinskii, V.~Vuleti{\'c}, and I.~L.
  Chuang, \enquote{Technologies for trapped-ion quantum information systems,}
  {\protect\JournalTitle{Quantum Information Processing}} \textbf{15},
  5351--5383 (2016).

\bibitem{Marquez2016}
A.~M{\'a}rquez~Seco, H.~Takahashi, and M.~Keller, \enquote{Novel ion trap
  design for strong ion-cavity coupling,} {\protect\JournalTitle{Atoms}}
  \textbf{4}, 15 (2016).

\bibitem{Ballance2017}
T.~Ballance, H.~Meyer, P.~Kobel, K.~Ott, J.~Reichel, and M.~K{\"o}hl,
  \enquote{Cavity-induced backaction in {P}urcell-enhanced photon emission of a
  single ion in an ultraviolet fiber cavity,} {\protect\JournalTitle{Phys. Rev.
  A}} \textbf{95}, 033812 (2017).

\bibitem{Takahashi2020}
H.~Takahashi, E.~Kassa, C.~Christoforou, and M.~Keller, \enquote{Strong
  coupling of a single ion to an optical cavity,}
  {\protect\JournalTitle{Physical Review Letters}} \textbf{124}, 013602 (2020).

\bibitem{Kimble1998}
H.~J. Kimble, \enquote{Strong interactions of single atoms and photons in
  cavity {QED},} {\protect\JournalTitle{Physica Scripta}} \textbf{1998}, 127
  (1998).

\bibitem{Harlander2010}
M.~Harlander, M.~Brownnutt, W.~H{\"a}nsel, and R.~Blatt, \enquote{Trapped-ion
  probing of light-induced charging effects on dielectrics,}
  {\protect\JournalTitle{New J. Phys.}} \textbf{12}, 093035 (2010).

\bibitem{Wang2011}
S.~X. Wang, G.~Hao~Low, N.~S. Lachenmyer, Y.~Ge, P.~F. Herskind, and I.~L.
  Chuang, \enquote{Laser-induced charging of microfabricated ion traps,}
  {\protect\JournalTitle{J. Appl. Phys.}} \textbf{110}, 104901 (2011).

\bibitem{Duan2004}
L.-M. Duan and H.~Kimble, \enquote{Scalable photonic quantum computation
  through cavity-assisted interactions,} {\protect\JournalTitle{Phys. Rev.
  Lett.}} \textbf{92}, 127902 (2004).

\bibitem{Hacker2016}
B.~Hacker, S.~Welte, G.~Rempe, and S.~Ritter, \enquote{A photon--photon quantum
  gate based on a single atom in an optical resonator,}
  {\protect\JournalTitle{Nature}} \textbf{536}, 193--196 (2016).

\bibitem{Pinotsi2008}
D.~Pinotsi and A.~Imamoglu, \enquote{Single photon absorption by a single
  quantum emitter,} {\protect\JournalTitle{Phys. Rev. Lett.}} \textbf{100},
  093603 (2008).

\bibitem{Lin2009}
G.~Lin, X.~Zou, X.~Lin, and G.~Guo, \enquote{Heralded quantum memory for
  single-photon polarization qubits,} {\protect\JournalTitle{Europhys. Lett.}}
  \textbf{86}, 30006 (2009).

\bibitem{Koshino2010}
K.~Koshino, S.~Ishizaka, and Y.~Nakamura, \enquote{Deterministic photon-photon
  {SWAP} gate using a {$\Lambda$} system,} {\protect\JournalTitle{Phys. Rev.
  A}} \textbf{82}, 010301 (2010).

\bibitem{Gea-Banacloche2011}
J.~Gea-Banacloche and L.~M. Pedrotti, \enquote{Comparative model study of
  two-photon deterministic passive quantum logical gates,}
  {\protect\JournalTitle{Phys. Rev. A}} \textbf{83}, 042333 (2011).

\bibitem{Rosenblum2011}
S.~Rosenblum, S.~Parkins, and B.~Dayan, \enquote{Photon routing in cavity
  {QED}: Beyond the fundamental limit of photon blockade,}
  {\protect\JournalTitle{Phys. Rev. A}} \textbf{84}, 033854 (2011).

\bibitem{Gea-Banacloche2012}
J.~Gea-Banacloche and L.~M. Pedrotti, \enquote{Single-photon, cavity-mediated
  gates: Detuning, losses, and nonadiabatic effects,}
  {\protect\JournalTitle{Phys. Rev. A}} \textbf{86}, 052311 (2012).

\bibitem{Rosenblum2017}
S.~Rosenblum, A.~Borne, and B.~Dayan, \enquote{Analysis of deterministic
  swapping of photonic and atomic states through single-photon {R}aman
  interaction,} {\protect\JournalTitle{Phys. Rev. A}} \textbf{95}, 033814
  (2017).

\bibitem{Bechler2018}
O.~Bechler, A.~Borne, S.~Rosenblum, G.~Guendelman, O.~E. Mor, M.~Netser,
  T.~Ohana, Z.~Aqua, N.~Drucker, R.~Finkelstein, Y.~Lovsky, R.~Bruch,
  D.~Gurovich, E.~Shafir, and D.~Dayan, \enquote{A passive photon-atom qubit
  swap operation,} {\protect\JournalTitle{Nat. Phys.}} \textbf{14}, 996--1000
  (2018).

\bibitem{Purcell1946}
E.~Purcell, \enquote{Spontaneous emission probabilities at radio frequencies,}
  {\protect\JournalTitle{Phys. Rev.}} \textbf{69}, 681 (1946).

\bibitem{Reiserer2015}
A.~Reiserer and G.~Rempe, \enquote{Cavity-based quantum networks with single
  atoms and optical photons,} {\protect\JournalTitle{Reviews of Modern
  Physics}} \textbf{87}, 1379--1418 (2015).

\bibitem{witthaut2010}
D.~Witthaut and A.~S. S{\o}rensen, \enquote{Photon scattering by a three-level
  emitter in a one-dimensional waveguide,} {\protect\JournalTitle{New Journal
  of Physics}} \textbf{12}, 043052 (2010).

\bibitem{aqua2019}
Z.~Aqua, M.~Kim, and B.~Dayan, \enquote{Generation of optical fock and w states
  with single-atom-based bright quantum scissors,}
  {\protect\JournalTitle{Photonics Research}} \textbf{7}, A45--A55 (2019).

\bibitem{Shomroni2014}
I.~Shomroni, S.~Rosenblum, Y.~Lovsky, O.~Bechler, G.~Guendelman, and B.~Dayan,
  \enquote{All-optical routing of single photons by a one-atom switch
  controlled by a single photon,} {\protect\JournalTitle{Science}}
  \textbf{345}, 903--906 (2014).

\bibitem{Rosenblum2016}
S.~Rosenblum, O.~Bechler, I.~Shomroni, Y.~Lovsky, G.~Guendelman, and B.~Dayan,
  \enquote{Extraction of a single photon from an optical pulse,}
  {\protect\JournalTitle{Nat. Photon.}} \textbf{10}, 19--22 (2016).

\bibitem{Carmichael1993}
H.~Carmichael, \enquote{Quantum trajectory theory for cascaded open systems,}
  {\protect\JournalTitle{Phys. Rev. Lett.}} \textbf{70}, 2273--2276 (1993).

\bibitem{Gardiner1985}
C.~Gardiner and M.~Collett, \enquote{Input and output in damped quantum
  systems: Quantum stochastic differential equations and the master equation,}
  {\protect\JournalTitle{Phys. Rev. A}} \textbf{31}, 3761--3774 (1985).

\bibitem{Gulati2017}
G.~K. Gulati, H.~Takahashi, N.~Podoliak, P.~Horak, and M.~Keller,
  \enquote{Fiber cavities with integrated mode matching optics,}
  {\protect\JournalTitle{Sci. Rep.}} \textbf{7}, 5556 (2017).

\bibitem{Takahashi2014}
H.~Takahashi, J.~Morphew, F.~Oru{\v{c}}evi{\'c}, A.~Noguchi, E.~Kassa, and
  M.~Keller, \enquote{Novel laser machining of optical fibers for long cavities
  with low birefringence,} {\protect\JournalTitle{Opt. Expr.}} \textbf{22},
  31317--31328 (2014).

\bibitem{Ott2016}
K.~Ott, S.~Garcia, R.~Kohlhaas, K.~Sch{\"u}ppert, P.~Rosenbusch, R.~Long, and
  J.~Reichel, \enquote{Millimeter-long fiber fabry-perot cavities,}
  {\protect\JournalTitle{Optics express}} \textbf{24}, 9839--9853 (2016).

\bibitem{Tokunaga2015}
Y.~Tokunaga and K.~Koshino, \enquote{A photon-photon controlled-phase gate
  using a {$\Lambda$} system,} in \emph{European Quantum Electronics
  Conference,}  (Optical Society of America, 2015), p. EB\_P\_7.

\end{thebibliography}

\end{document}